\documentstyle[11pt,aaspp4]{article}
\lefthead{TELFER ET AL.}
\righthead{BAL SYSTEM OF QSO SBS1542+541}

\newcommand{\kms}{${\rm km \ s^{-1}}$}
\newcommand{\col}{${\rm cm^{-2}}$}
\newcommand{\flux}{${\rm erg \ cm^{-2} \ s^{-1} \ \AA ^{-1}}$} 
\newcommand{\nuflux}{${\rm erg \ cm^{-2} \ s^{-1} \ Hz^{-1}}$}
\newcommand{\lam}{$\lambda$}
\newcommand{\lamlam}{$\lambda \lambda$}
\newcommand{\lya}{Ly$\alpha$}
\newcommand{\captionsize}{\footnotesize}
\newcommand{\captionbaseline}{\baselineskip 0.14in}
\newcommand{\plotland}[1]{\plotfiddle{#1}{2.9in}{-90}{40}{40}{-155}{230}}

\begin{document}
\title{The Very Highly Ionized Broad Absorption Line
System of the QSO SBS1542+541}
\author{Randal C. Telfer, Gerard A. Kriss, Wei Zheng, and Arthur F. Davidsen}
\affil{Center for Astrophysical Sciences, Johns Hopkins University, Baltimore,
MD, 21218-2686; rt19@pha.jhu.edu, gak@pha.jhu.edu, zheng@pha.jhu.edu, 
afd@pha.jhu.edu}
\and
\author{Richard F. Green}
\affil{Kitt Peak National Observatory, National Optical Astronomy
Observatories, P.O. Box 26732, \\
950 North Cherry Ave., Tucson, AZ, 85726-6732}

\begin{abstract}
We have analyzed the broad absorption line system of the bright ($V\sim 16.5$)
high-redshift ($z=2.361$)
QSO SBS1542+541 using UV spectra from the HST FOS along with optical data
from the MMT and the Steward Observatory 2.3m telescope.  These
spectra offer continuous wavelength coverage from
$1200$ to ${\rm 8000 \ \AA}$,
corresponding to $\sim {\rm 340-2480 \ \AA}$ in the QSO rest frame.  The
line of sight to the object contains only three identified intervening 
Lyman-limit absorption
systems.  Only one of these is optically
thick at the Lyman edge, a low-redshift ($z=0.156$) system with a strong
Lyman edge observed at ${\rm 1055 \ \AA}$ (${\rm 314 \ \AA}$ in the rest 
frame) in a Hopkins Ultraviolet Telescope spectrum from the Astro-2 mission.  
The spectra therefore offer a rare opportunity to study broad absorption lines 
in the rest-frame extreme UV.

We find that the broad absorption line system is lacking in species of 
relatively low ionization often seen in broad absorption systems, such as 
\ion{C}{3}, \ion{O}{3}, and \ion{Si}{4}.  Instead, the
system is dominated by very high-ionization species.  The
strongest features correspond to \ion{O}{6}, \ion{Ne}{8}, and 
\ion{Si}{12}.  In addition to other high-ionization lines, we identify
apparently saturated broad Lyman-series lines of order Ly$\gamma$ and higher.

There is strong evidence for partial occultation of the QSO emission source,
particularly from the higher-order Lyman lines which indicate a 
covered fraction less than $0.2$.  With the exception of \ion{C}{4} and
\ion{N}{5}, which are low-ionization species in the context of this 
system, all of the other lines depress the flux by more than 20\%.  Absorption 
from \lya\ also depresses the flux more than 20\%, indicating that there are 
at least two different regions contributing to \ion{H}{1} absorption.  
Overall, the data suggest a correlation between a larger covered fraction 
and a higher state of ionization.  These observations reveal inhomogeneity in
the ionization structure of the broad absorption line gas.

We have used photoionization models to constrain the total column density
and ionization state of the system.  A single-slab model
consistent with our observational limits on the column densities requires
$N_H \approx 5 \times 10^{22}$ \col\ and an incident ionization parameter 
$U \approx 2$.  Since the observed covered fractions suggest multiple
zones, we also produced a two-slab model and find 
$10^{21} \ {\rm cm}^{-2} < N_H < 10^{23} \ {\rm cm}^{-2}$ 
and $0.08 < U < 4$ for the smaller zone, and 
$N_H \gtrsim 3 \times 10^{21}$ \col\ and $U \gtrsim 2$ for the larger zone.
We suggest that the different covered fractions can be 
explained by either a special line of sight through a disk-like geometry or by
the existence of density fluctuations of a factor $\gtrsim 2$ in the BAL gas.  
The large column density and high state of ionization suggest 
that the system is likely associated with an X-ray ``warm absorber''.
\end{abstract}

\section{INTRODUCTION}

Observations of SBS1542+541 using the International Ultraviolet Explorer
(Porter {\it et al.}\ unpublished\footnote{The research was submitted in 1991 
for publication in {\it ApJ}, but the work was never 
fully completed due to the illness of Alain C. 
Porter that led to his untimely death in 1993.})
show that there is significant continuum flux down to below ${\rm 1200 \ \AA}$,
corresponding to $\sim {\rm 360 \ \AA}$ in
the rest frame.  This made the object a good candidate for possibly 
observing \ion{He}{2} \lya\
\lam 304 absorption in the intergalactic medium with the Hopkins
Ultraviolet Telescope (HUT) during the Astro-2 mission.  Observations
of HS1700+64 ($z=2.743$) on the same mission successfully measured the 
\ion{He}{2} opacity (Davidsen, Kriss, \& Zheng 1996\markcite{dkz96}).  
However, during the 
observations of SBS1542+541 on 1995 March 16 we discovered that there was a 
cutoff in flux slightly
redshifted from the expected position of the \ion{He}{2} \lya\ absorption,
causing us to suspect that the observed feature was a Lyman limit from
a low-redshift intervening absorber.  To further investigate this
possibility we performed observations on 1995 September 3 with
the G130H, G190H, and G270H gratings of the HST FOS.  The presence of the
intervening system is verified by a strong \ion{Mg}{2}
\lamlam 2796, 2804 doublet in the G270H spectrum at the same 
redshift ($z=0.156$) as the
Lyman edge visible in the HUT spectrum.  Fortuitously, the FOS
spectra also reveal broad absorption lines from several very highly-ionized
species, many of which have never before been observed in a broad
absorption line QSO.  It is the study of this broad absorption system 
that is the focus of our present interest.

Roughly 10\% of radio-quiet QSOs exhibit broad ($\gtrsim 2000$ \kms) 
blueshifted absorption features associated with the QSO, known as
broad absorption lines (BALs),
with outflow velocities generally $0-0.1c$ (Turnshek 1984\markcite{tu84}).
These objects are collectively referred to as broad absorption line QSOs,
or BALQSOs.  The absorption features most commonly observed are \lya\ 
\lam 1216, \ion{C}{4} \lam 1549, \ion{Si}{4} \lam 1397, and \ion{N}{5}
\lam 1240, though low-ionization lines, usually 
\ion{Mg}{2} \lam 2798 and \ion{Al}{3} \lam 1857, are observed in some
objects, usually called low-ionization BALQSOs.  Higher-ionization 
lines are sometimes reported, up through \ion{Ne}{8} \lam 774 
and \ion{Mg}{10} \lam 615 (Korista \& Arav 1997\markcite{koar92}).

When studying a BAL system, one of the primary goals is to determine
column densities of the various observed ions, which are used in turn
to derive the ionization state of the gas and its chemical
composition.  To determine column densities, it is necessary to
measure the optical depth of the absorption as a function of outflow 
velocity.  In deriving the optical depth from the relative depression of
the continuum flux, one typically makes the assumptions that (1) there are no 
unresolved narrow lines in the BAL troughs, (2) scattered radiation does
not fill in the absorption troughs, and (3) the BAL gas completely
and homogeneously
covers the continuum source on the observed line of sight.  If 
any of these assumptions are invalid, then the true column
densities could be much greater than those inferred from the observations.
Based on high-resolution observations of BAL troughs 
(Barlow \& Junkkarinen 1994\markcite{baju94}),
assumption (1) appears to be a good one.  Studies of BAL geometry
(Turnshek 1988\markcite{tu88}; 
Hamann, Korista, \& Morris 1993\markcite{hcm93}) suggest that
the global BAL region covering factor, which is the fractional solid angle 
subtended by the BAL gas as seen from the continuum source, is generally 
$\lesssim 0.2$ (except possibly in low-ionization BALQSOs; 
Boroson \& Meyers 1992\markcite{bome92}; 
Turnshek {\it et al.}\ 1997\markcite{tuea97}).  If the global BAL covering 
factor is small, and especially if the outflow velocities are large, 
then (2) should
be adequately satisfied, although spectropolarimetry suggests that some 
scattered light is present in the depths of the BAL troughs 
(Goodrich \& Miller 1995\markcite{gomi95}; 
Cohen {\it et al.}\ 1995\markcite{coea95}).  However, there is evidence that 
assumption (3) is invalid:  in some instances absorption doublets appear to 
show saturation of the components but not at zero intensity
(Korista {\it et al.}\ 1992\markcite{koea92}), implying that the absorbing gas
only partially covers the source.  More definite evidence for partial
covering has been observed in the so-called associated or
$z_{abs}\approx z_{em}$ absorption lines 
(Hamann {\it et al.}\ 1997\markcite{hbjb97}; 
Hamann, Barlow, \& Junkkarinen 1997\markcite{hbj97}),
which are distinguished from BALs primarily by being much narrower, usually
having widths $\lesssim$ a few hundred \kms.  The fraction of the 
continuum source occulted by the BAL region along the line of sight is
called the covered fraction, or for clarity the {\em continuum source 
covered fraction}, to distinguish it from the global BAL region covering
factor.  For the remainder of this paper, 
we will use the term {\em covered fraction} to refer to the continuum source 
covered fraction.

We begin with a presentation of the observations used for this research 
(\S\ref{sec:obs}), followed by a discussion of the spectra and 
how we fit the BALs in order to estimate minimum column densities and 
covered fractions of the absorbing ions (\S\ref{sec:analysis}).
We then discuss some properties of the system that can be inferred in the 
context of a photoionization model, including constraints on the ionization 
state and column density and the resulting implications for the geometry and
physical structure of the gas (\S\ref{sec:discuss}).  Finally we summarize the
characteristics of this particular BALQSO that make it so interesting and
potentially relevant to the study of BALQSOs as a whole (\S\ref{sec:summary}).

\section{OBSERVATIONS\label{sec:obs}}

\begin{figure}
\plotland{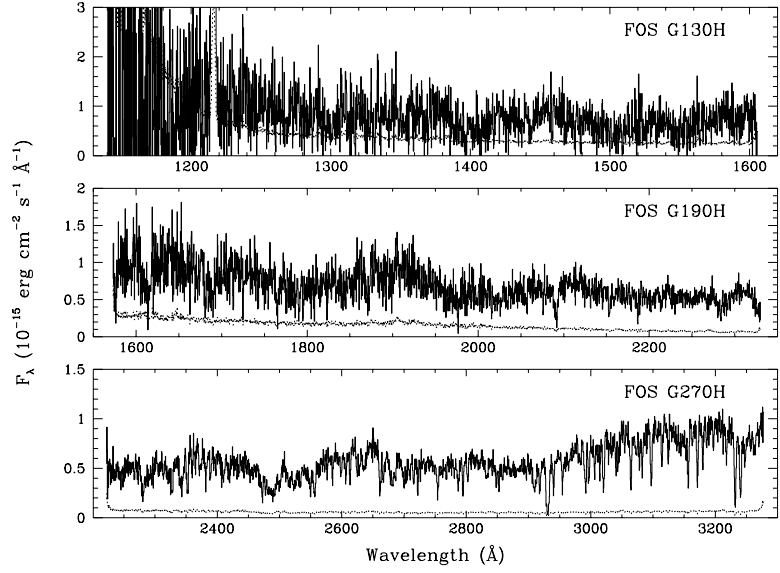}
\caption{\protect\captionsize \protect\captionbaseline
HST FOS spectra of SBS1542+541 (solid lines)
along with the 1$\sigma$ errors per pixel (dotted lines).}
\label{fig:uvspec}
\end{figure}

Our observations with the HST FOS were performed on 1995 September 3.  They 
include 14880~s with the G130H grating, 5470~s with G190H, and 1600~s with 
G270H.  These three spectra are shown in Figure~\ref{fig:uvspec}.  The
G130H data had to be corrected for a flickering diode not
recognized by the FOS pipeline at the time of processing.  This diode 
(\#22 affecting wavelengths $1162-1167 \ {\rm \AA}$) was 
added to the list of dead diodes and the spectrum was reprocessed.  These 
spectra have a resolution of 1 \AA\ (G130H), 1.44 \AA\ (G190H), and 
2 \AA\ (G270H), corresponding to a range of velocity resolution of
$183-287$ \kms\ for the FOS data.

\begin{figure}
\plotland{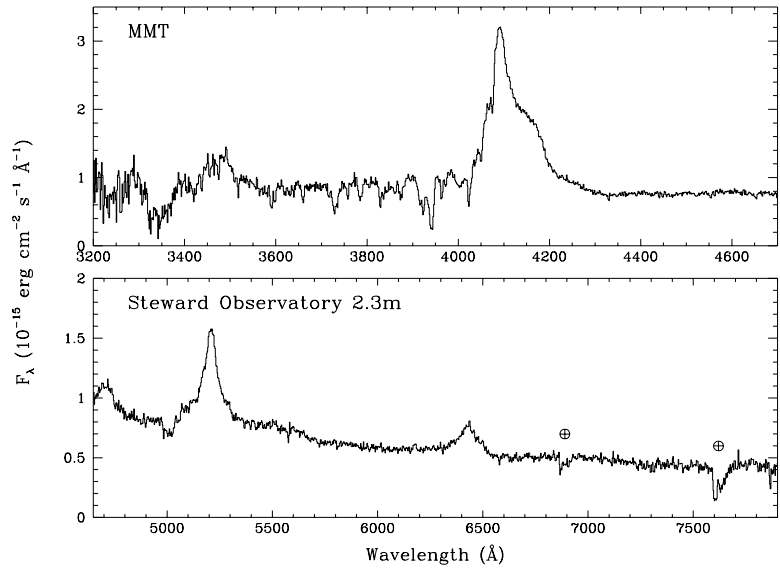}
\caption{\protect\captionsize \protect\captionbaseline
Optical spectra of SBS1542+541 from the MMT
and Steward Observatory 2.3m telescope.}
\label{fig:optspec}
\end{figure}

At optical wavelengths we use two observations.  The first was performed on
1995 April 1 with the MMT and covers ${\rm 3120-4770 \ \AA}$ with a 
resolution of 
3.2 \AA, corresponding to a velocity resolution $200-307$ \kms.  The other 
observation was performed on 1995 June 3 with the Steward Observatory 2.3m 
telescope, including the wavelength range ${\rm 4640-8010 \ \AA}$ with a 
5.54 \AA\ resolution, corresponding to
a velocity resolution of $207-358$ \kms.  The MMT spectrum is of excellent 
quality in the wavelength region below ${\rm 3500 \ \AA}$ for a ground-based 
observation, which is fortunate since this reveals well the broad \ion{O}{6} 
absorption feature centered at $\sim {\rm 3350 \ \AA}$.  The optical spectra 
are shown in Figure~\ref{fig:optspec}.

\begin{figure}
\plotland{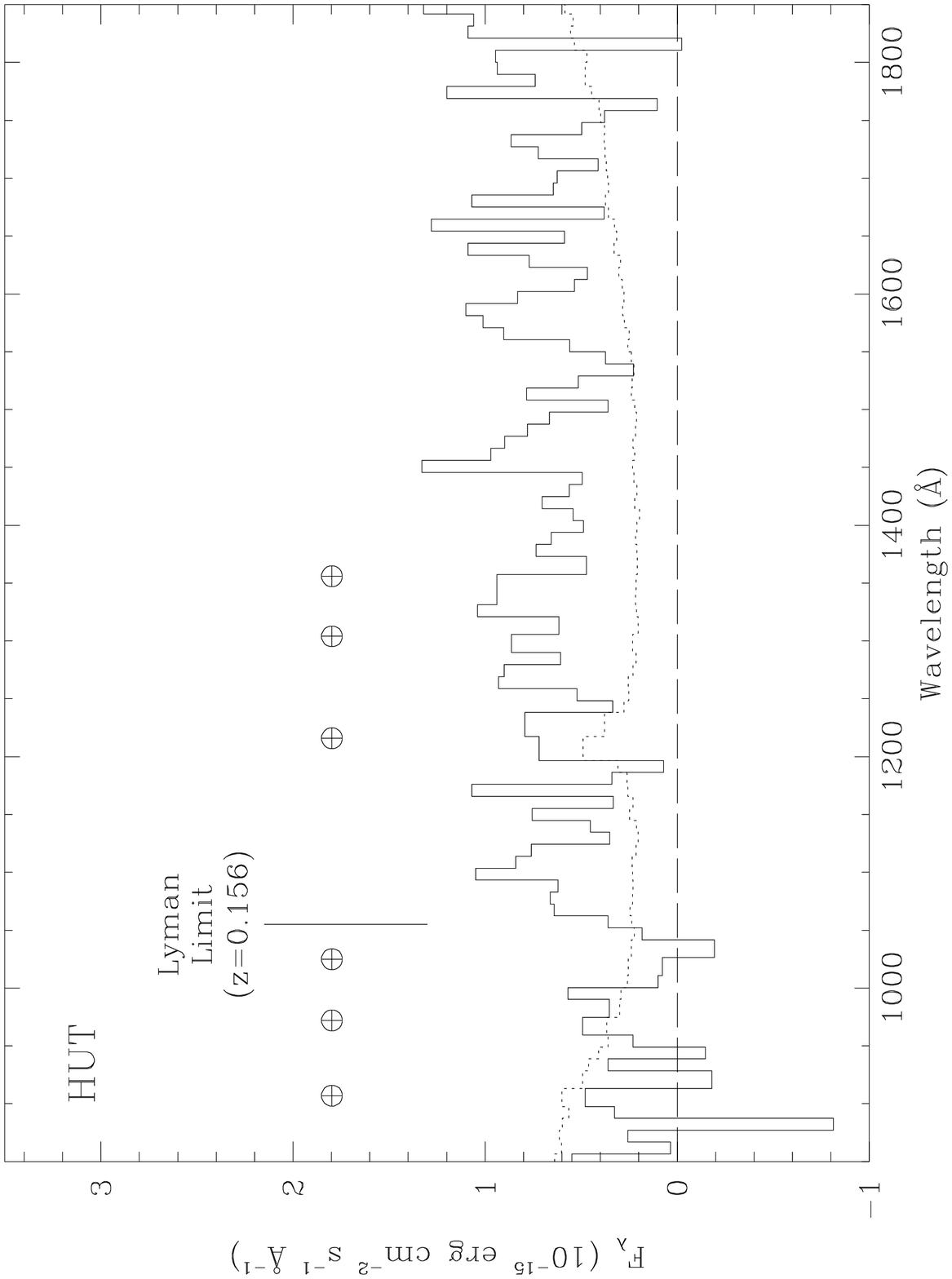}
\caption{\protect\captionsize \protect\captionbaseline
HUT spectrum of SBS1542+541 (solid line)
after airglow subtraction, 
rebinned by 10 \AA, along with 1$\sigma$ errors per bin (dotted line).
The \earth\ symbols indicate where pixels were removed to eliminate the 
residual airglow from the strongest lines that remained after airglow
subtraction using a blank-field template.
The location of the Lyman-limit feature from the $z=0.156$ intervening 
system is marked.}
\label{fig:hutspec}
\end{figure}

Together, these five spectra provide us with continuous wavelength coverage 
from $1200$ to ${\rm 8000 \ \AA}$, or $\sim {\rm 340-2480 \ \AA}$ in the QSO 
rest frame.  For the purpose of identifying the $z=0.156$ intervening
absorption system, we also refer to the HUT spectrum of SBS1542+541 obtained
on 1995 March 16 during
the Astro-2 mission, shown in Figure~\ref{fig:hutspec}.  A normalized template 
created from blank-field spectra was used to remove airglow.  This removes
the broad wings of \lya, but some narrow residuals remain for \lya\ and
a few other strong lines which were removed by hand.
The spectrum is of low S/N due to the short integration time (4320 s),
so we do not use it for the analysis of the BAL system.

We have also analyzed X-ray data from a $5632 \ {\rm s}$ exposure 
of SBS1542+541 obtained with the
{\it ROSAT} Position Sensitive Proportional Counter (PSPC) 
on 1993 August 14.  The observed count rate was
$7.02 \times 10^{-3} \ {\rm count \ s}^{-1}$.  The S/N is too
low to derive meaningful information about the spectral shape, so we
use the data only to estimate the X-ray flux.  This is discussed further
in \S\ref{sec:photoion}.

\section{ANALYSIS\label{sec:analysis}}
\subsection{General Comments}

We corrected the spectra for Galactic extinction based on a
column density of neutral hydrogen along the line of sight of
$N(H) = 1.30 \times 10^{20}$ \col\ 
(Stark {\it et al.}\ 1992\markcite{stae92}).  Using a 
conversion of $E_{(B-V)}=N(H) / 5.2 \times 10^{21} \ {\rm cm^{-2}}$ 
(Shull \& Van Steenberg 1985\markcite{shva85}), this
corresponds to an extinction $E_{(B-V)} = 0.025$.  We then performed the 
extinction correction using the formulation of
Cardelli, Clayton, \& Mathis (1989\markcite{ccm89}), assuming $R_V = 3.1$.

In order to estimate the errors in the optical data, we fit small bands
of apparently featureless continuum in each spectrum.  Near the
\ion{C}{4} \lamlam 1548, 1551 feature in the Steward data, we fit the region
${\rm 4840-4955 \ \AA}$.  We find
that the error per pixel is around $2.8 \times 10^{-17}$ \flux, 
corresponding to a S/N per pixel of $\sim 30$.  We used this
value to produce a constant error array that was used in the fit of
\ion{C}{4}.  In the MMT spectrum, we fit 
${\rm 3667-3707 \ \AA}$ and found an error per pixel of
$4.5 \times 10^{-17}$ \flux, corresponding to a S/N per pixel of
$\sim 20$.  Again we simply produced a constant error array that was used
in the fit of \lya, \ion{N}{5} \lamlam 1239, 1243, and \ion{O}{6} 
\lamlam 1032, 1038.  However, given that the 
observation includes short wavelengths so near the atmospheric limit, 
the S/N is lower at the shorter wavelengths.
Hence, we realize that
this is probably an overestimate of the errors in the region
of \lya\ and \ion{N}{5} and an underestimate of the errors near
\ion{O}{6}, but since there are no spectral regions near these lines that
can be well fit to estimate the errors, we will use this constant value.

Many of the problems associated with studying BALs are
minimized in the spectrum of SBS1542+541.  Due to the relatively small 
velocity width of the lines as well as the apparent lack of low-ionization 
species, most of the interesting lines are not blended and therefore the 
continuum is generally well-defined in the vicinity of the broad absorption 
features.  Also, the line of sight to the object contains an
unusual paucity of intervening Lyman-limit absorption systems. We identify 
only three at $z=2.246$, $z=0.721$, and $z=0.156$.  In the next
section we briefly discuss each narrow-line system.  A more comprehensive
study of the intervening narrow-line systems ({\it e.g.}\ Vogel \& Reimers 
1995\markcite{vore95}) would require data of higher S/N.

\subsubsection{Intervening Narrow-Line Systems}

The $z=2.246$ system is 
clearly established by definite Lyman-series lines through 
Ly$\epsilon$ as well as lines from Ly$6-8$ that appear to be blended with
other narrow lines, presumably from the \lya\ forest.  In addition, there are
strong lines corresponding to \ion{C}{3} \lam 977 and \ion{C}{2} \lam 903
along with possible lines from \ion{O}{4} \lam 788 and \ion{O}{3} \lam 833.
This system overlaps the broad absorption in velocity space, 
precluding an accurate estimate of the column density from the Lyman-edge
opacity, since the relative contribution of each system to the flux depression
around ${\rm 2950 \ \AA}$ is not known.  However, given that we find in
\S\ref{sec:cover} that most of 
the neutral hydrogen from the BAL system covers only $\sim$20\% of the 
continuum source, we can conclude that at least some of this flux depression
must be due to the Lyman limit of the $z=2.246$ system, since the full
depression is $\sim$40\% of the total continuum flux.  This enables us
to estimate that the total column of neutral hydrogen in this system 
probably lies in the range $\log N(H) \approx 16.65-16.90$.  The system could
be intrinsic rather than intervening; however, given that the observed metal 
lines suggest a lower ionization level than the broad lines, we assume in
our analysis that it is physically distinct from the highly-ionized BAL system.
 
The system at $z=0.721$ is identified mainly on the basis of the very strong
lines of \lya\ ($W_\lambda \approx {\rm 2.8 \ \AA}$) and Ly$\beta$ 
($W_\lambda \approx {\rm 1.8 \ \AA}$) that have unusually broad profiles, the 
\lya\ line having a Full Width at Zero Absorption (FWZA) $\approx$900 \kms\ 
and Ly$\beta$ having a FWZA$\approx$750 \kms.  A weak Ly$\gamma$ feature can 
be seen, as well as a possible
feature from \ion{C}{3} \lam 977 blended with the broad absorption of
\ion{Si}{12} \lam 521.  It is difficult to confirm the presence of any
Lyman-limit feature, which should lie at ${\rm 1569 \ \AA}$, due to the 
uncertain continuum shape in this region.
There does appear to be some small decrease in the flux, but the situation
is complicated by the fact that QSOs can only rarely be
observed so far into the rest-frame UV, and consequently knowledge of the
broad emission line properties in this region is extremely limited. 
We estimate $\log N(H) \lesssim 16.8$.

The redshift of the $z=0.156$ system is calculated from the strong
\ion{Mg}{2} \lamlam 2796, 2804 doublet that occurs around
${\rm 3235 \ \AA}$.  Absorption from \lya\ and possibly \ion{C}{4}
\lamlam 1548, 1551 are also detected.  None of the higher-order
Lyman lines can be seen since they lie below the wavelength coverage
of our data.  However, the system is firmly
established by the \ion{Mg}{2} doublet and by the presence of the strong
Lyman-limit feature at $\sim {\rm 1055 \ \AA}$ visible in the HUT spectrum 
of this object (Figure~\ref{fig:hutspec}).

\subsection{Broad Absorption Lines}

We have identified broad absorption features corresponding to resonance lines
of \ion{H}{1}, \ion{C}{4}, \ion{N}{5}, \ion{O}{6}, \ion{Ne}{5} through 
\ion{Ne}{8}, \ion{Mg}{10}, \ion{Si}{12}, and \ion{Ar}{7}.  Each feature 
is primarily characterized by a single trough that is blue-shifted at 
line center by $\sim 11000-12000$ \kms\ with 
respect to the rest frame and has a width of $2000-3000$ \kms.  
The troughs are fairly flat across the bottom with no obvious consistent
structure and
have sharp velocity cutoffs on both the red and blue end.  Because of the 
high ionization levels and small widths of the lines, many 
of the doublets are resolved in velocity space, including
\ion{Ne}{8} \lamlam 770, 780, \ion{Mg}{10} \lamlam 610, 625, 
and \ion{Si}{12} \lamlam 499, 521.  The Lyman-series lines through Ly$\epsilon$
are also resolved, with the exception of Ly$\beta$ which is strongly blended 
with \ion{O}{6} \lamlam 1032, 1038.

There are several other possible lines that are either part of a strong
blend or occur in portions of the spectrum where the continuum level is
not well-defined and hence the existence of the line is difficult to
determine:
\begin{enumerate}
\item{Absorption from \ion{O}{4} \lam 788 could be present around 
${\rm 2550 \ \AA}$, but the continuum shape is uncertain due to the 
\ion{Ne}{8} broad emission line and a blend of narrow absorption lines.}
\item{There appears to be some slight evidence for
an \ion{O}{5} \lam 630 feature around ${\rm 2035 \ \AA}$, but the absorption 
is either weak or the covered fraction is small ($\lesssim 0.2$).  Also, this 
lies in between the apparent \ion{O}{5} broad emission line and the 
\ion{Mg}{10} \lam 625 BAL, again making the continuum shape uncertain.}
\item{There is a dip in the flux shortward of an apparent 
broad emission line around ${\rm 1900 \ \AA}$, possibly 
\ion{Ne}{6} \lam 559 + \ion{Ne}{5} \lam 568.  The dip extends from 
$\sim {\rm 1670 \ \AA}$ to $\sim {\rm 1830 \ \AA}$ and could be a blend of 
several possible BALs including \ion{O}{4} \lamlam 553, 554, \ion{Ne}{5} 
\lam 568, \ion{Ne}{6} \lam 559, and \ion{Al}{11} \lamlam 550, 568.}
\end{enumerate}
For the reasons stated above and the uncertainties in the profiles,
we make no attempt to fit or deblend any of the above lines.
We did measure the equivalent width of the broad feature around 
${\rm 1400 \ \AA}$ 
since this feature appears to be well-defined.  This is likely a blend
of \ion{Ne}{6} \lam 433, \ion{Mg}{8} \lam 430, and possibly 
\ion{Mg}{7} \lam 429.  Table~\ref{ta:atomicdata} lists atomic data for all 
of the resonance lines relevant to this paper.

\begin{deluxetable}{lccrl}
\tablecaption{Atomic Data\tablenotemark{a}\label{ta:atomicdata}}
\tablehead{			&	\colhead{Creation}&
\colhead{Destruction}		&&
\colhead{Oscillator}\\
\colhead{Ion}			&	\colhead{IP(eV)}&	
\colhead{IP(eV)}	&	\colhead{\lam (\AA)}	&
\colhead{Strength}}
\startdata
\ion{H}{1} 	&   0.0 &  13.6 & 1215.67 & 0.416\nl
		&       &       & 1025.72 & 0.0790\nl
		&       &       &  972.54 & 0.0290\nl
		&       &       &  949.74 & 0.0139\nl
		&       &       &  937.80 & 0.00780\nl
\ion{C}{4}	&  47.9 &  64.5 & 1548.20 & 0.190\nl
		&       &	& 1550.77 & 0.0952\nl
\ion{N}{5}	&  77.5	&  97.9 & 1238.82 & 0.157\nl
		&	&	& 1242.80 & 0.0780\nl
\ion{O}{5}	&  77.4	& 113.9 &  629.73 & 0.515\nl
\ion{O}{6} 	& 113.9	& 138.1 & 1031.93 & 0.133\nl
		&	&	& 1037.62 & 0.0661\nl
\ion{Ne}{5}	&  97.1 & 126.2 &  480.41 & 0.110\nl
		&	&	&  568.42 & 0.0928\nl
\ion{Ne}{6}	& 126.2 & 157.9 &  399.82 & 0.0844\nl
		&	&	&  401.14 & 0.168\nl
		&	&	&  433.18 & 0.0505\nl
		&	&	&  558.59 & 0.0907\nl
\ion{Ne}{7}	& 157.9	& 207.3	&  465.22 & 0.389\nl
\ion{Ne}{8}	& 207.3	& 239.1	&  770.41 & 0.103\nl
		&	&	&  780.32 & 0.0505\nl
\ion{Mg}{7}	& 186.5 & 225.0 &  429.13 & 0.0801\nl
\ion{Mg}{8}	& 225.0 & 266.0	&  430.47 & 0.0763\nl
\ion{Mg}{10}	& 328.2 & 367.5 &  609.79 & 0.0842\nl
		&	&	&  624.95 & 0.0410\nl
\ion{Al}{11}	& 399.4 & 442.1 &  550.03 & 0.0773\nl
		&	&	&  568.15 & 0.0375\nl
\ion{Si}{4}	&  33.5 &  45.1 & 1393.76 & 0.524\nl
		&	&	& 1402.77 & 0.260\nl 
\ion{Si}{12}	& 476.1 & 523.5	&  499.40 & 0.0719\nl
		&	&	&  520.67 & 0.0345\nl
\ion{Ar}{7}	&  91.0 & 124.3 &  585.75 & 1.24\nl
\tablenotetext{a}{From Verner, Barthel, \& Tytler (1994\markcite{vbt94})}
\enddata
\end{deluxetable}

Although the absorption system of SBS1542+541 qualifies it as a BALQSO
according to the canonical definition of the class 
(Weymann, Carswell, \& Smith 1981\markcite{wcs81}), it certainly has 
characteristics that distinguish it from ``classical'' BALQSOs.
The line widths of $2000-3000$ \kms\ are relatively small---absorption over
a range of velocities exceeding $10000$ \kms\ is common in BALQSOs.  In 
addition, the features
from \ion{C}{4}, \ion{N}{5}, and \lya\ are quite weak relative to 
``classical'' systems, and absorption
from \ion{Si}{4} \lamlam 1394, 1403, a feature that is virtually omnipresent 
in BAL systems, is absent completely.  Thus objects such as SBS1542+541 
would be excluded from typical searches for BALQSOs which look for these 
features to be strong.  We also find no evidence for absorption by other 
commonly observed low-ionization species such as \ion{C}{3}, \ion{N}{3},
\ion{N}{4}, and \ion{O}{3}.  Absorption from \ion{P}{5} \lamlam 1118, 1128,
sometimes reported in broad absorption, is also not observed.  There is a 
feature with the right profile shape
and at the right wavelength to be \ion{Fe}{3} \lam 1123, but the feature is
weak and the existence of \ion{Fe}{3} would require a state of ionization
much lower than that inferred from the other lines (\S\ref{sec:photoion}), 
so we dismiss the feature as a probable cluster of \lya -forest lines.
The absence of low-ionization species, particularly
\ion{Si}{4}, is useful for constraining the ionization state of
the gas (\S\ref{sec:discuss}).

\begin{figure}
\plotland{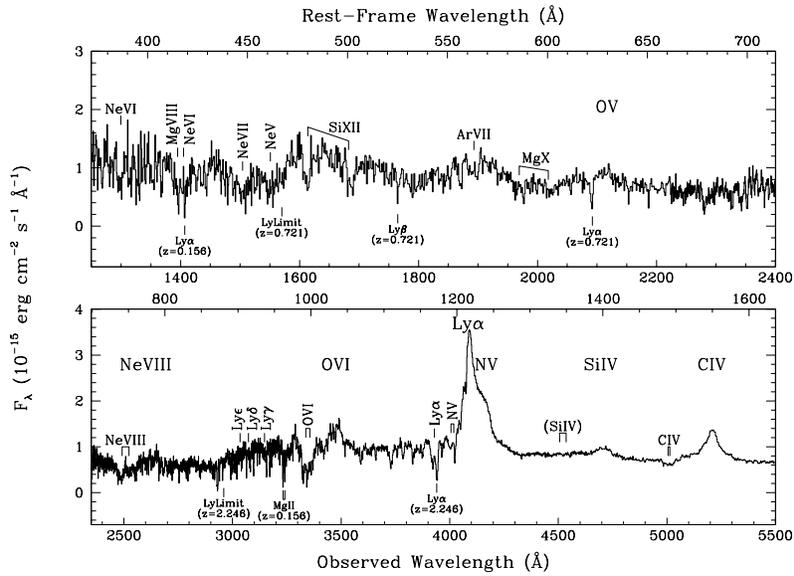}
\caption{\protect\captionsize \protect\captionbaseline
Combined spectra, extinction-corrected and rebinned for higher S/N.
The broad absorption lines and prominent 
broad emission lines are labeled above the spectrum, some intervening 
absorption features below.}
\label{fig:speclabel}
\end{figure}

In Figure~\ref{fig:speclabel} we show for reference a combined spectrum of 
the FOS and optical data, corrected for Galactic extinction and rebinned 
for higher S/N.  All of the identified broad absorption features are labeled, 
as are the prominent broad emission lines and several intervening absorption 
features.

\subsection{Broad Line Fits\label{sec:linefit}}

For fitting purposes we attempted to develop optical depth templates 
from various observed absorption profiles such as
\ion{C}{4} \lamlam 1548, 1551 and \ion{Ne}{8} \lamlam 770, 780
(see Korista {\it et al.}\ 1992\markcite{koea92}).  However,
we found that no single derived optical depth template was able to fit all of 
the observed lines satisfactorily, due mainly to slight variations in the
widths of the profiles.  Instead of a rigid template, it is necessary to use
a more flexible absorption profile.  We use two such profiles, one a 
simple Gaussian and the other a profile that we will refer
to as the ``flat profile'', an absorption profile that we synthesized in order
to match the primary characteristics of the observed troughs, specifically
the relatively flat bottoms and sharp velocity cutoffs.  The flat profile
is a feature of constant optical depth smoothed at each edge by a 
semi-Gaussian with FWHM fixed at the instrumental resolution. 
Each of these profiles provides a simple way to characterize the absorption 
features by three free parameters:  the velocity centroid, the FWHM, and the 
equivalent width.

Table~\ref{ta:flatfitparams} shows the resulting fit parameters for the
flat template, and Table~\ref{ta:gaussfitparams} shows the parameters for the
Gaussian fits.  All of the fits were performed with the IRAF task
{\em specfit} (Kriss 1994\markcite{kr94}).  We fit
a separate profile to each doublet component, even when the components are 
strongly blended as is the case for \ion{C}{4} and \ion{N}{5}.  We constrained
the velocity centroids and widths of the individual components of the 
multiplet to be the same during the fits.  Whenever 
two lines are blended, the equivalent width of the entire blend
is given.  The quoted errors on all quantities are purely $1\sigma$
confidence limits derived from the error matrix of the fit.

\begin{deluxetable}{lcccc}
\tablecaption{Fit Parameters:  Flat Profile\label{ta:flatfitparams}}
\tablehead{
\colhead{Ion}		&	\colhead{\lam (\AA)}	&
\colhead{centroid(\kms)}	&	\colhead{FWHM(\kms)}	&
\colhead{$W_{\lambda}$(\AA)}}
\startdata
\ion{H}{1} 	& 1215.67 & $-11480\pm 30$ & $2860\pm 70$ & $10.24\pm 0.65$\nl
		&  972.54 & $-11120\pm 20$ & $1940\pm 50$ & $3.68\pm 0.30$\nl
		&  949.74 & & & $3.08\pm 0.31$\nl
		&  937.80 & & & $3.78\pm 0.30$\nl
\ion{C}{4}	& 1548.20, 1550.77& $-11400\pm 70$ & $3190\pm 190$ & 
	$7.95\pm 0.80$\nl
\ion{N}{5}	& 1238.82, 1242.80& $-11370\pm 50$ & $27400\pm 100$ & 
	$5.34\pm 0.51$\nl
\ion{O}{6} 	& 1031.93, 1037.62& $-11340\pm 30$ & $3260\pm 70$ &
	$33.5\pm 1.6$\nl
\ion{Ne}{5}	&  480.41 & $-11810\pm 70$ & $2970\pm 150$ & $4.84\pm 0.79$\nl
\ion{Ne}{6}	& 399.82, 401.14 & $-10410\pm 50$ & $2530\pm 120$ & 
	$5.97\pm 0.80$\nl
\ion{Ne}{7}	&  465.22 & $-11350\pm 80$ & $3150\pm 180$ & $6.00\pm 0.88$\nl
\ion{Ne}{8}	&  770.41 & $-11360\pm 20$ & $2730\pm 40$ & $10.07\pm 0.51$\nl
		&  780.32 &                &              & $6.05\pm 0.37$\nl
\ion{Mg}{10}	&  609.79 & $-11750\pm 50$ & $3110\pm 110$& $4.50\pm 0.82$\nl
		&  624.95 &                &              & $3.91\pm 0.70$\nl
\ion{Si}{12}	&  499.40 & $-11570\pm 60$ & $3160\pm 130$& $5.64\pm 0.76$\nl
		&  520.67 &                &              & $5.86\pm 0.67$\nl
\ion{Ar}{7}	&  585.75 & $-11540\pm 60$ & $2910\pm 130$ & $4.34\pm 0.73$\nl
\ion{Ne}{6}+\ion{Mg}{8} & 433.18, 430.47 &\nodata & \nodata & $11.2\pm 1.1$\nl
\enddata
\end{deluxetable}

\begin{deluxetable}{lcccc}
\tablecaption{Fit Parameters:  Gaussian\label{ta:gaussfitparams}}
\tablehead{
\colhead{Ion}		&	\colhead{\lam (\AA)}	&
\colhead{centroid(\kms)}	&	\colhead{FWHM(\kms)}	&
\colhead{$W_{\lambda}$(\AA)}}
\startdata
\ion{H}{1} 	& 1215.67 & $-11290\pm 90$ & $2150\pm 100$ & $9.95\pm 0.60$\nl
		&  972.54 & $-11360\pm 50$ & $1480\pm 100$ & $3.42\pm 0.30$\nl
		&  949.74 & & & $2.81\pm 0.29$\nl
		&  937.80 & & & $3.02\pm 0.29$\nl
\ion{C}{4}	& 1548.20, 1550.77& $-11330\pm 100$ & $3110\pm 330$ & 
	$10.8\pm 1.8$\nl
\ion{N}{5}	& 1238.82, 1242.80& $-11570\pm 190$ & $2160\pm 200$ & 
	$7.70\pm 0.50$\nl
\ion{O}{6} 	& 1031.93, 1037.62& $-11090\pm 160$ & $3010\pm 230$ &
	$29.9\pm 3.2$\nl
\ion{Ne}{5}	&  480.41 & $-11950\pm 280$ & $2850\pm 610$ & $5.6\pm 1.1$\nl
\ion{Ne}{6}\tablenotemark{a} & 399.82, 401.14 & \nodata & \nodata & \nodata \nl
\ion{Ne}{7}	&  465.22 & $-11530\pm 240$ & $2830\pm 620$ & $7.0\pm 1.3$\nl
\ion{Ne}{8}	&  770.41 & $-11630\pm 80$ & $2530\pm 180$ & $12.24\pm 0.57$\nl
		&  780.32 &                &               & $7.33\pm 0.52$\nl
\ion{Mg}{10}	&  609.79 & $-11670\pm 220$ & $3050\pm 540$ & $5.1\pm 1.3$\nl
		&  624.95 &                 &               & $4.7\pm 1.1$\nl
\ion{Si}{12}	&  499.40 & $-11200\pm 580$ & $2730\pm 270$ & $6.5\pm 1.1$\nl
		&  520.67 &                 &               & $7.46\pm 0.64$\nl
\ion{Ar}{7}	&  585.75 & $-11500\pm 280$ & $2760\pm 410$ & $5.31\pm 0.67$\nl
\ion{Ne}{6}+\ion{Mg}{8} & 433.18, 430.47 &\nodata & \nodata & $14.5\pm 3.2$\nl
\tablenotetext{a}{Parameters for the \ion{Ne}{6} feature are not listed 
because no minimum in $\chi ^2$ could be achieved that provided a good fit 
to the data}
\enddata
\end{deluxetable}

\begin{figure}
\plotland{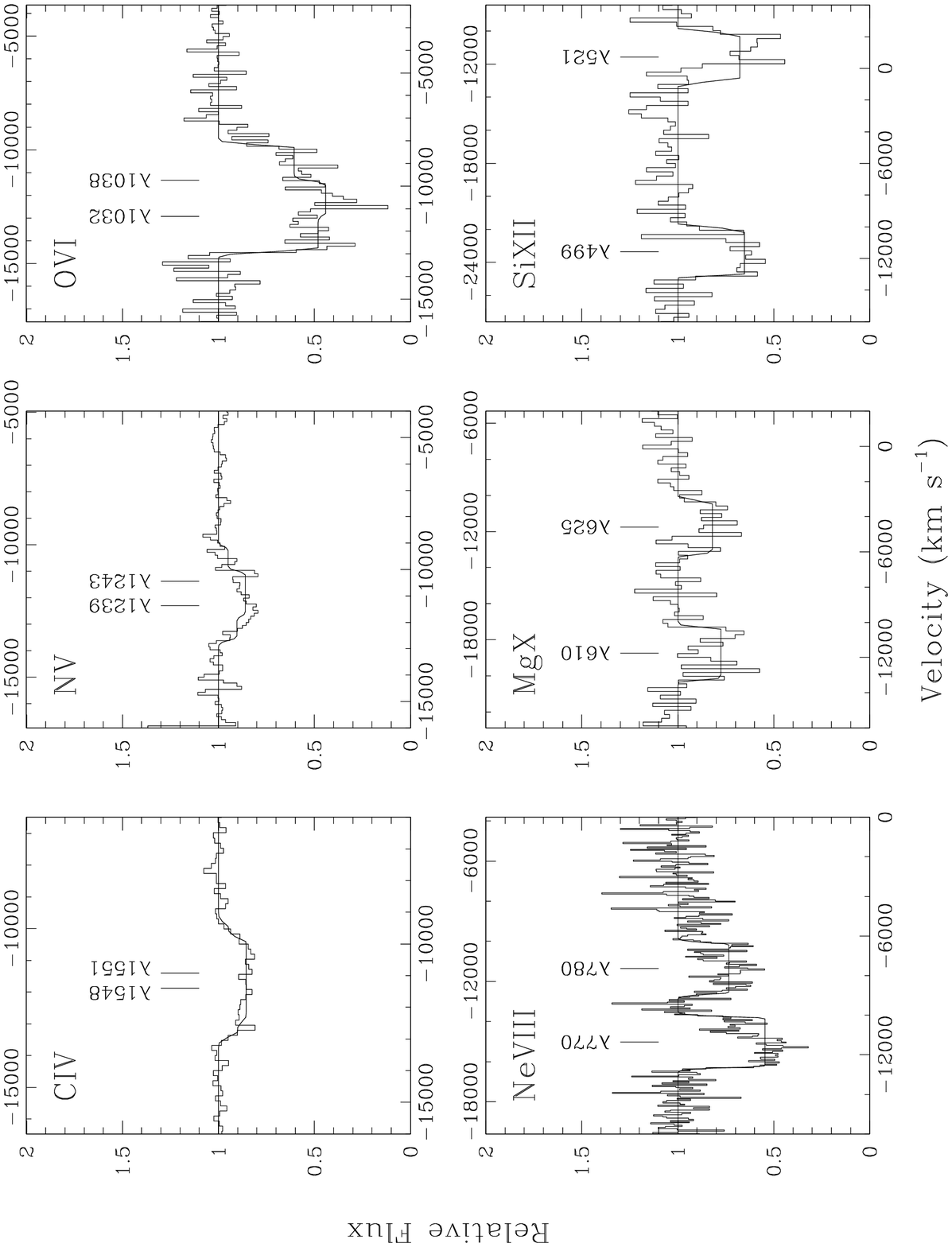}
\caption{\protect\captionsize \protect\captionbaseline
Absorption profiles of observed 
doublets from lithium-like ions along with the best-fit flat profiles.
The identified narrow lines and the known Ly$\beta$ in O VI have been removed.
The fits for the strongly-blended doublets of C IV and N V assume a covered 
fraction of unity, while the other fits allow for a partial covered fraction.
The spectra are plotted as flux relative to 
the best-fit continuum against velocity relative to the wavelength of the 
blue (bottom axis) and red (top axis) components of
the doublet in the QSO rest frame.  For display purposes, 
the FOS G190H spectrum, which includes Mg X and Si XII, is rebinned by four 
pixels for higher S/N.}
\label{fig:profile}
\end{figure}

Only the most obvious narrow lines
(FWHM$\sim$instrumental resolution, significance$\gtrsim 3\sigma$) were 
deblended from the broad lines, with all other features in the wavelength 
range of the BALs being treated as part of the BAL profiles.  Less-significant 
narrow lines were sometimes fit outside the BAL profiles to improve the 
continuum fit.  The presence of other narrow lines in the 
BAL profiles could cause additional uncertainty in the fit parameters.  
This is particularly true
for \lya, \ion{N}{5}, \ion{O}{6}, and Ly$\gamma$ through Ly$\epsilon$,
which lie in the portion of the spectrum where the \lya\ forest is most
dense.  The fits for \ion{Si}{12} and to a lesser extent \ion{Mg}{10} 
show some deviation from the observed profiles (see Figure~\ref{fig:profile}) 
indicating that these also are probably somewhat corrupted by narrow lines.  
The possible effects of blending on the derived column densities and 
covered fractions are discussed in \S\ref{sec:colden} and \S\ref{sec:cover},
respectively.
For the fit of \ion{O}{6}, the Ly$\beta$ BAL corresponding to the best fit of
the higher-order Lyman lines, including a partial covered fraction as 
discussed in \S\ref{sec:cover}, was also deblended from the absorption profile.
If there is additional Ly$\beta$ corresponding to the stronger \lya\ line,
we could still be overestimating the strength of \ion{O}{6}.

Both profiles are capable of producing reasonable fits to the data, which is 
why we fit every feature with each profile.  Because we created the flat 
profile specifically to fit the shapes of the observed absorption features, 
it achieves more satisfying fits than the Gaussian by eye, 
but only slightly better in $\chi ^2$ because of the generally low S/N.  The 
tails of the Gaussians extend out slightly beyond where there is any apparent 
absorption, causing the equivalent widths of the Gaussian fits to be generally 
somewhat larger than those of the flat profile.  Because the flat profile 
provides better fits, we believe that it parameterizes better the basic 
characteristics of the lines.  However, using both profiles helps to 
characterize the uncertainties of our results associated with the uncertainty 
of the profile shape.

\subsection{Minimum Column Densities\label{sec:colden}}

As we discuss in \S\ref{sec:cover}, we believe that the
covered fractions of the different absorbing ions are significantly less
than unity, and that the absorption may in many cases be
optically thick.  Without knowing the covered fraction, it is
not possible to determine accurately the optical depth of the lines, and 
hence to determine the column densities.  We can, however, put a good 
lower limit on the absorbing columns by calculating the column densities under
the assumption of a covered fraction of unity.  Rather than estimate the
columns from the linear portion of the curve of growth, we derive values
from direct integration of the optical depth of the best-fit absorption 
profiles.  We produce the best fit possible for the lines from each ion 
assuming a covered fraction of unity, with the constraint that the ratio 
of the optical depths of the individual components is fixed by the oscillator 
strengths.  To facilitate this, for the Gaussian
fits we use a Gaussian defined in terms of optical depth rather than flux. 
The column densities with statistical errors as calculated from these fits 
are shown in Table~\ref{ta:colden}.  Since none of the lines are optically 
thick under the assumption of total covering, the uncertainty in the 
measured column densities due to the possible existence of narrow-line 
blends in the BALs should be relatively small.

\begin{deluxetable}{lcc}
\tablecaption{Minimum Column Densities\label{ta:colden}}
\tablehead{
&	\colhead{Flat Profile} & \colhead{Gaussian}\\
\colhead{Ion}		 &	\colhead{log $N_j$(\col)}&	
\colhead{log $N_j$(\col)}}
\startdata
\ion{H}{1}\tablenotemark{a} 	& $14.84\pm 0.03$ & $14.89\pm 0.05$\nl
\ion{C}{4}	& $14.64\pm 0.04$ & $14.77\pm 0.06$\nl
\ion{N}{5}	& $14.74\pm 0.04$ & $14.84\pm 0.04$\nl
\ion{O}{6} 	& $15.77\pm 0.04$ & $15.79\pm 0.04$\nl
\ion{Ne}{5}	& $15.91\pm 0.08$ & $15.89\pm 0.14$\nl
\ion{Ne}{6}     & $15.83\pm 0.08$ & \nodata \nl 
\ion{Ne}{7}	& $15.51\pm 0.08$ & $15.39\pm 0.14$\nl
\ion{Ne}{8}	& $15.90\pm 0.02$ & $15.98\pm 0.03$\nl
\ion{Mg}{10}	& $15.87\pm 0.06$ & $15.92\pm 0.10$\nl
\ion{Si}{12}	& $16.25\pm 0.06$ & $16.34\pm 0.08$\nl
\ion{Ar}{7}	& $14.62\pm 0.08$ & $14.70\pm 0.10$\nl
\tablenotetext{a}{Derived from \lya\ only}
\enddata
\end{deluxetable}

Some of the ions, particularly \ion{Mg}{10}, \ion{Si}{12}, and \ion{H}{1}, 
exhibit multiple lines with similar apparent optical depths despite
different oscillator strengths, suggesting that the lines are likely
saturated and therefore that the continuum source is not fully occulted
by the absorber.  As a result, the fits to
these features under the assumption of a covered fraction of unity are
poor, particularly for the higher-order Lyman lines, for which no reasonable 
fit could be achieved.  Also, none of the BAL troughs
depress the flux more than $\sim 60$\%, consistent with partial covering.  
In the next section we discuss the necessity of including partial covered
fractions for fitting and interpreting the BAL system.

\subsection{Partial Covered Fractions\label{sec:cover}}

\begin{figure}
\plotland{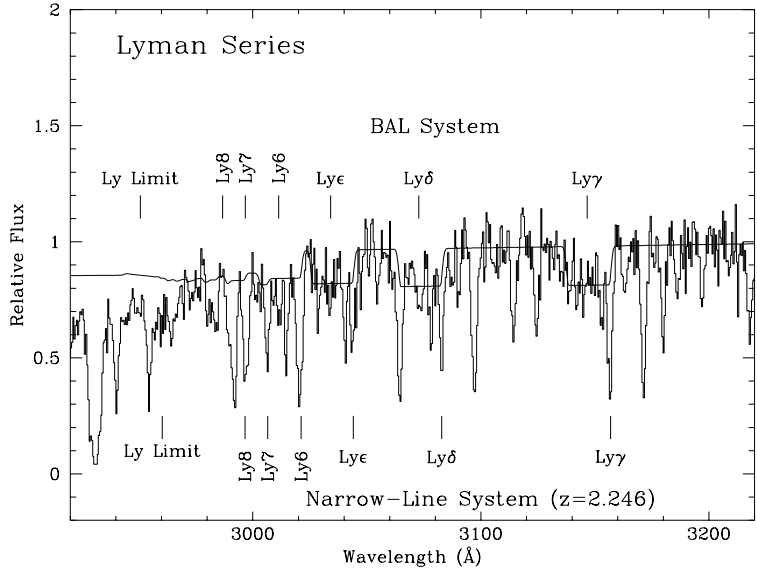}
\caption{\protect\captionsize \protect\captionbaseline
Best fit of the broad Lyman-series absorption lines 
of order Ly$\gamma$ and higher, plotted as flux relative to the best-fit
continuum.  To emphasize the broad lines the contribution 
of the narrow lines to the fit are not shown.  Around ${\rm 2980 \ \AA}$ 
the depression
of the continuum flux due to the $z=2.246$ narrow-line system 
becomes significant.  The location of the Lyman limit and Lyman series lines 
from this narrow-line system are indicated.  The narrow lines of order Ly6 and 
higher are strongly blended with other lines.}
\label{fig:lyman}
\end{figure}

In principle,
given a fully-resolved absorption multiplet, one can uniquely determine
both the optical depth and covered fraction as a function of velocity
(Hamann {\it et al.}\ 1997\markcite{hbjb97}).  However, given that the 
quality of the
data does not permit such a detailed analysis, we proceed under the 
assumption that the covered fraction is a constant function of velocity, but
may vary from ion to ion.  Allowing the covered fraction to vary, we 
re-fit the doublets of \ion{O}{6}, \ion{Ne}{8}, \ion{Mg}{10} and 
\ion{Si}{12}.  Figure~\ref{fig:profile} shows the absorption profiles and 
best-fit flat profiles of these ions with a partial covered fraction after 
the removal of the narrow lines and the known Ly$\beta$ in \ion{O}{6}.  
The fits for the strongly-blended doublets of the other identified 
lithium-like ions \ion{C}{4} and \ion{N}{5} are also shown
for completeness, assuming a covered fraction of unity.  The best fit of the 
higher-order Lyman lines, showing only the 
contributions from the continuum and broad absorption features, is plotted in 
Figure~\ref{fig:lyman}.  We not only fit the resolved features Ly$\gamma$ 
through Ly$\epsilon$ but also included lines through Ly29 to fit the spectrum 
down to the BAL Lyman limit.  Around $2980 \ {\rm \AA}$ 
the continuum flux begins to be
depressed significantly beyond the 20\% level, presumably due to the high-order
Lyman-line and Lyman-limit absorption from the $z=2.246$ narrow-line
system.

\begin{figure}
\plotland{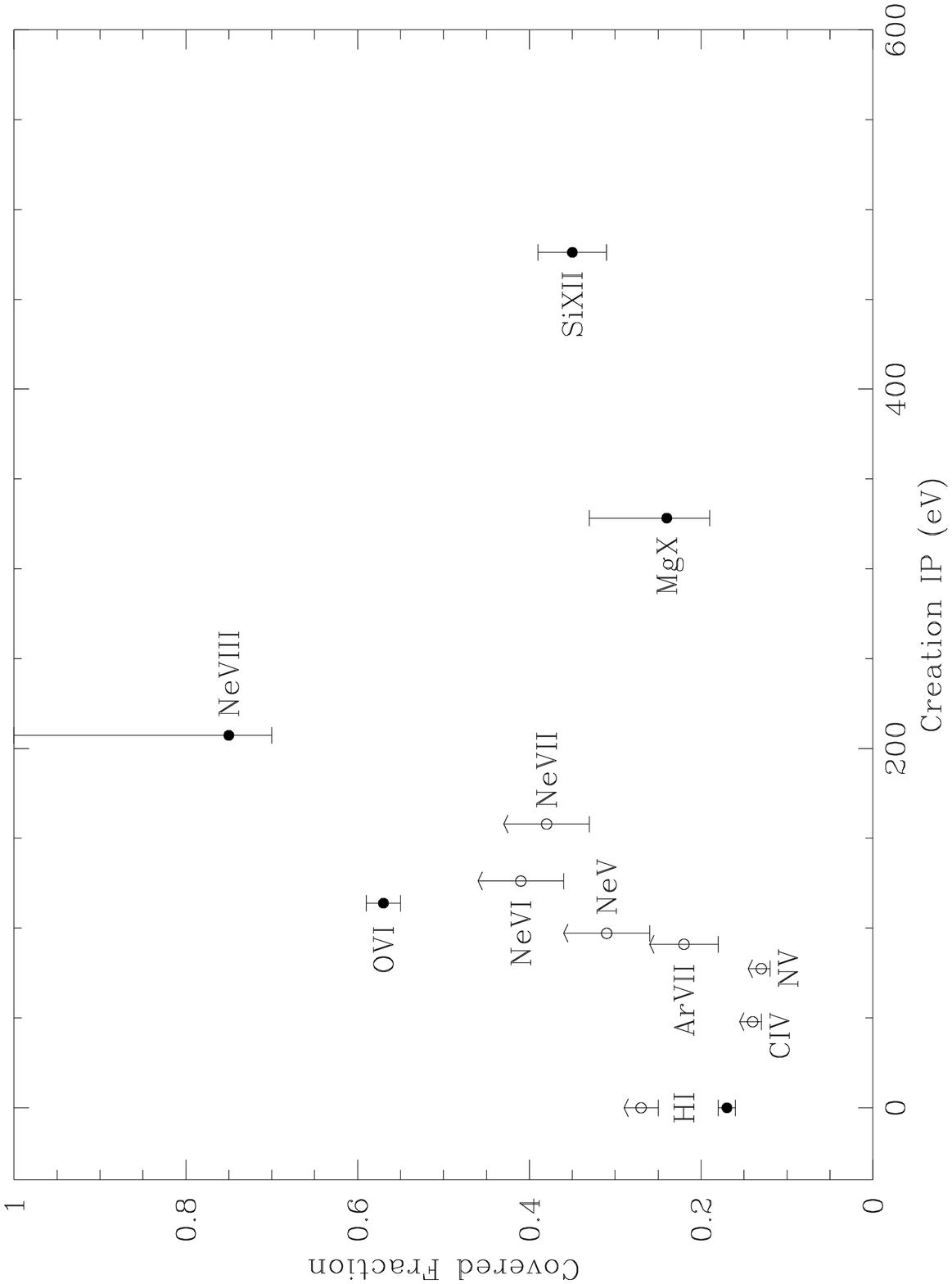}
\caption{\protect\captionsize \protect\captionbaseline
Covered fractions and creation IPs of various ions
derived from the flat profile fits.  The closed circles are
best-fit values derived from multiple components, while the open circles 
represent lower limits based on the penetration depths of the 
troughs.  The errors, particularly on the best-fit values, could
be significantly larger if the observed BALS are blended with other features.
However, the best-fit covered fraction of H I is fairly reliable since it
is determined from several lines.  Two values are plotted for H I, the minimum 
value derived from \lya\ only and the best-fit value based on the higher-order 
Lyman lines.}
\label{fig:cover}
\end{figure}

The resulting best-fit values for the covered fractions are listed in
Table~\ref{ta:cover}, along with minimum covered fractions for the 
singlet features derived from the penetration depths of the troughs.
As is evident from comparing the values derived from the different
absorption profiles, the uncertainty in constraining the 
covered fractions due to the choice of absorption profile is larger in most 
cases than that derived from the statistics of the fits.  
In Figure~\ref{fig:cover} we plot the covered fractions from the 
flat profile fits against the 
creation ionization potential for each ion.  The closed circles are
best-fit values derived from multiple components, while the open circles 
represent lower limits.  It is possible that many or most of the singlet 
absorption features are saturated.  If this is the case, then the open
circles in Figure~\ref{fig:cover} closely represent the true covered fractions
and not just lower limits.

\begin{deluxetable}{lcc}
\tablecaption{Covered Fractions\label{ta:cover}}
\tablehead{
&	\colhead{Flat Profile} & \colhead{Gaussian}\\
\colhead{Ion}		 &	\colhead{$c_f$}&	
\colhead{$c_f$}}
\startdata
\ion{H}{1}\tablenotemark{a} 	& $>0.27\pm 0.02$ & $>0.34\pm 0.02$\nl
\ion{H}{1}\tablenotemark{b}	& $0.17\pm 0.01$  & $0.16\pm 0.01$\nl
\ion{C}{4}	& $>0.14\pm 0.01$ & $>0.18\pm 0.02$\nl
\ion{N}{5}	& $>0.13\pm 0.01$ & $>0.21\pm 0.03$\nl
\ion{O}{6} 	& $0.57 \pm 0.02$  & $0.52\pm 0.01$\nl
\ion{Ne}{5}	& $>0.31\pm 0.05$ & $>0.35\pm 0.07$\nl
\ion{Ne}{6}	& $>0.41\pm 0.05$ & \nodata \nl
\ion{Ne}{7}	& $>0.38\pm 0.05$ & $>0.40\pm 0.07$\nl
\ion{Ne}{8}     & $0.75^{+0.25}_{-0.05}$  & $0.81^{+0.19}_{-0.10}$\nl
\ion{Mg}{10}	& $0.24^{+0.09}_{-0.05}$  & $0.22\pm 0.03$\nl
\ion{Si}{12}	& $0.35\pm 0.04$  & $0.38\pm 0.03$\nl
\ion{Ar}{7}	& $>0.22\pm 0.04$ & $>0.28\pm 0.02$\nl
\tablenotetext{a}{Derived from \lya\ only}
\tablenotetext{b}{Derived from Lyman-series lines of order Ly$\gamma$ and
higher}
\enddata
\end{deluxetable}

Blending of the observed BALs with weak broad lines or with narrow lines 
could be important in the determination of the covered fractions because a 
fairly small change in the profile of one of the components of a doublet 
could cause a significant change in the inferred covered fraction.  As 
mentioned previously, it appears that such contamination does exist in
the \ion{Si}{12} and \ion{Mg}{10} doublets, and it is likely that this
is true to some extent for the \ion{O}{6} and \ion{Ne}{8} features as well.
Hence the true uncertainties in the covered fractions for these ions are
probably somewhat larger than the listed statistical errors.  This
is not so much the case for \ion{H}{1}, for which the covered fraction is
determined from several lines and is therefore more reliable.

Despite this complication, it is clear that the covered 
fraction for the dominant absorbing source is not the same for all ions.  
The most convincing evidence for this is provided by the higher-order Lyman 
lines, for
which we can reliably constrain the covered fraction to be less than 0.2.  
This covered fraction is less than that of most of the other lines, 
including the doublets of \ion{O}{6}, \ion{Ne}{8}, and \ion{Si}{12}, as well
as the observed singlet features, all of which depress the flux more that
20\%.  Therefore there must be multiple absorbing regions contributing to 
the BAL system, providing direct evidence for inhomogeneity in the ionization 
structure of the BAL gas.  There appears to be a general trend for the
very high-ionization species (creation IP $\gtrsim 100 \ {\rm eV}$) to occult 
more of the continuum source than the lower-ionization species 
(creation IP $\lesssim 100 \ {\rm eV}$).

Since we can constrain well the covered fraction of \ion{H}{1}, it is useful 
to estimate a better lower limit on the column density of \ion{H}{1},
as this provides a very useful constraint for our 
photoionization models (\S\ref{sec:discuss}).  For the fit of the Lyman
lines with a partial covered fraction, a large column density of
\ion{H}{1}, $\log N(H) > 17$, is required to obtain a good fit.
However, as mentioned previously, there is probably some narrow-line 
blending, which could cause us to slightly overestimate the
strength of the higher-order Lyman lines.  To get a more conservative lower
limit on the column density of \ion{H}{1}, we allowed for the contribution
of additional narrow lines below the $3\sigma$ significance level that 
appear to be blended with the broad lines.  With these lines included,
we find that with 90\% confidence a column density of 
$\log N(H)\gtrsim 16.87$ is necessary to provide a good fit using a 
flat profile, and $\log N(H)\gtrsim 16.83$ using a Gaussian profile.  
An important point is
that we can make these estimates because we can clearly see the higher-order
Lyman lines usually not visible in BAL spectra.  The fact that the \lya\ 
feature indicates a larger covered fraction than the higher-order Lyman 
lines suggests that there are at least two different regions causing 
absorption 
in \ion{H}{1} -- a region with a small covered fraction but a high column 
density of \ion{H}{1} that contributes the absorption for the higher-order 
Lyman lines, and a region with a lower column density of \ion{H}{1} but 
significantly higher covered fraction that dominates the observed \lya\ 
absorption profile.  As a result, if we were to
estimate the total column of \ion{H}{1} in this BAL system from \lya\ alone
and assume a covered fraction of unity, 
we would underestimate the true value by orders of magnitude.

\section{DISCUSSION\label{sec:discuss}}

\subsection{Photoionization Models\label{sec:photoion}}

\begin{figure}
\plotland{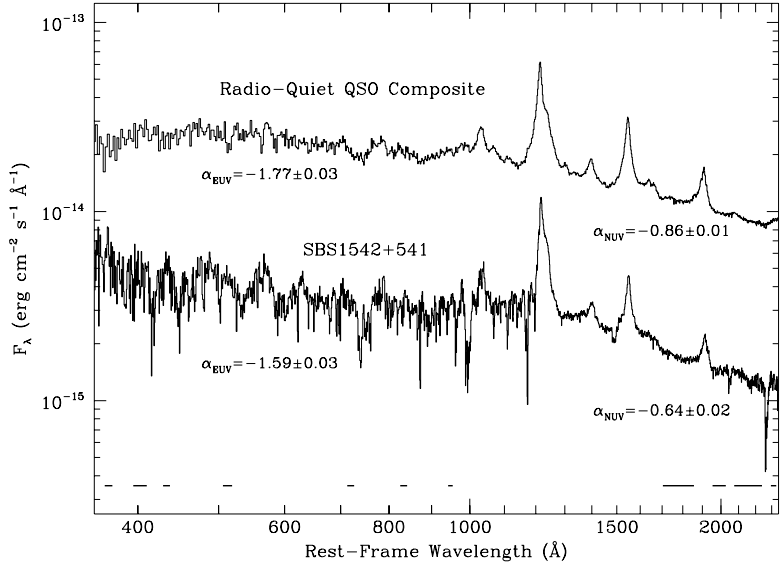}
\caption{\protect\captionsize \protect\captionbaseline
Combined spectra, corrected for intervening
absorption, shown with the radio-quiet QSO composite of Zheng {\it et al.}\ 
(1997) for comparison.  The composite has been arbitrarily normalized.  
The lines below the spectra indicate
the wavelength regions used to fit the continuum of SBS1542+541.  This 
continuum exhibits a break in the power-law index around 1000 \AA, similar 
to the composite spectrum.}
\label{fig:corspec}
\end{figure}

In order to produce self-consistent photoionization models we construct an
ionizing continuum based on the observed continuum.  However, to accurately 
represent the EUV continuum, particularly for a high-redshift QSO,
we need to first correct for the accumulated Lyman-line and 
Lyman-continuum absorption from intervening clouds, the primary characteristic
of which is a broad trough centered around ${\rm 700 \ \AA}$ known as the 
Lyman valley
(M\o ller \& Jakobsen 1990\markcite{moja90}).  We use the empirical formula
for the statistics of the forest-line distribution
\begin{equation}
\frac{\partial ^2}{\partial z \partial N} = A(1+z)^\gamma N^{-\beta}
\end{equation}
(Press \& Rybicki 1993\markcite{prry93}).  For the parameter values we adopt 
$\beta = 1.5$, $A=2.4 \times 10^7$, and $\gamma = 2.46$, since these
yield a line number in agreement with observations of high-redshift QSOs
(Hu {\it et al.}\ 1995\markcite{huea95}).  Using this distribution we applied
the statistical correction for $z=2.361$, including intervening 
absorbers with a total column density between $10^{13}$ \col\ and 
$10^{16}$ \col.  Any absorption system with a column density greater
than $\sim 10^{16}$ \col\ produces a significant Lyman limit feature
for which we can correct directly, which we did for the $z=0.721$ and
$z=2.246$ systems.  The resulting corrected spectrum is
plotted in Figure~\ref{fig:corspec}, shifted into the QSO rest frame.
Also shown for comparison, with arbitrary normalization, is
the composite spectrum of radio-quiet QSOs produced by Zheng
{\it et al.}\ (1997\markcite{zeea97}).  The lines below the spectrum indicate
the wavelength regions used to fit the continuum of SBS1542+541.  
The spectrum shows a break in the power-law index $\alpha$ 
($L_\nu \propto \nu^\alpha$) around ${\rm 1000 \ \AA}$.  Such a break is
consistent with Zheng {\it et al.}\ (1997\markcite{zeea97}), 
who find that a break in the UV power-law index around ${\rm 1050 \ \AA}$ is a 
general feature of QSO continua.
From the Steward spectrum we derive a near-UV power-law index
of $\alpha_{NUV}=-0.64\pm 0.02$ by fitting the region 
${\rm 1700-2240 \ \AA}$, longward of the break.  Shortward
of the break, we use the FOS data to derive an extreme-UV power-law 
index of $\alpha_{EUV}=-1.59\pm 0.03$ by fitting the continuum range
${\rm 360-960 \ \AA}$.

\begin{figure}
\plotland{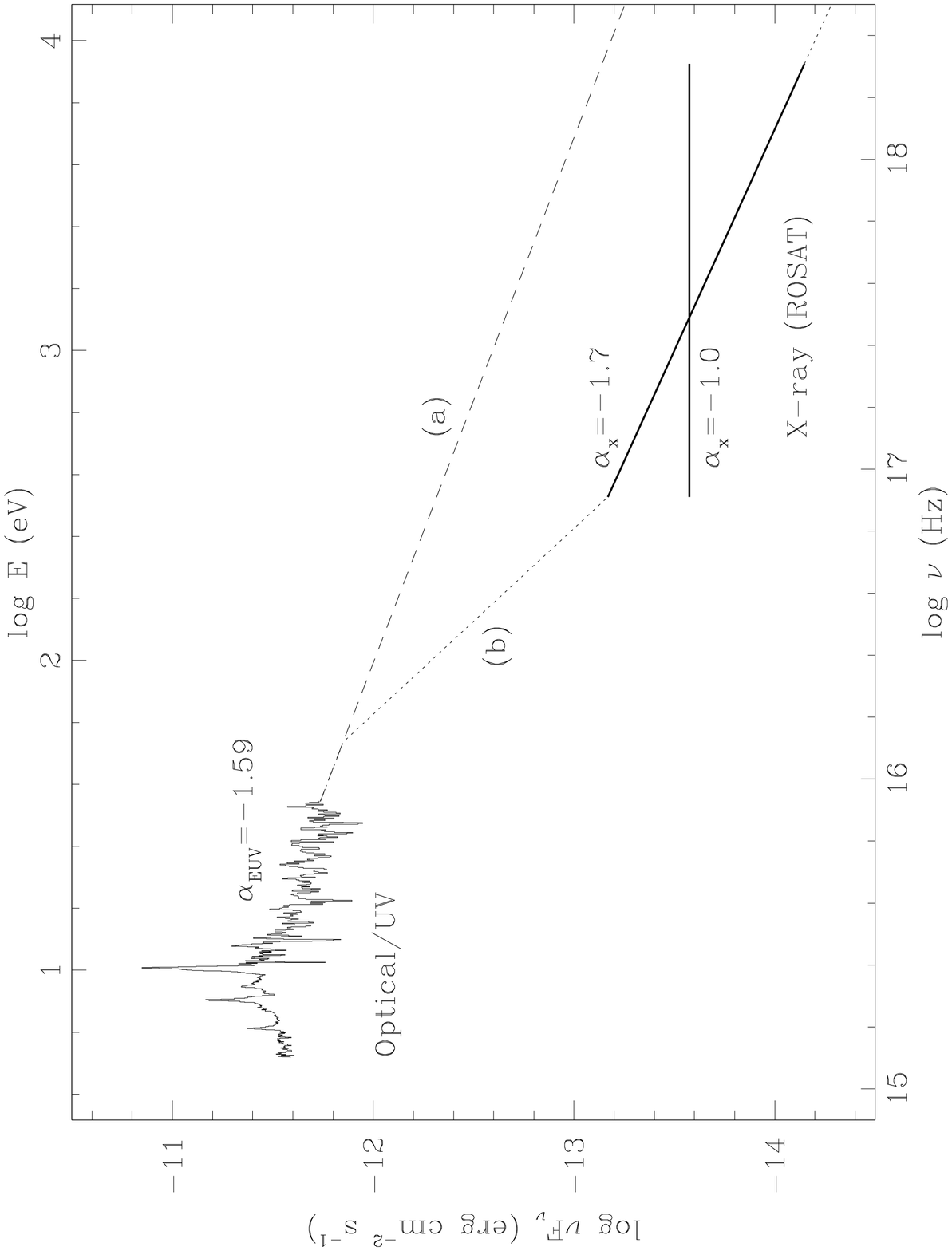}
\caption{\protect\captionsize \protect\captionbaseline
Broad-band continuum of SBS1542+541,
including the optical/UV spectra and the X-ray continuum as normalized 
by {\it ROSAT} data.  The two lines for the X-ray continuum correspond to
different assumptions for the X-ray spectral index, $\alpha_x=-1.7$
and $\alpha_x=-1.0$.  The dashed and dotted lines, (a) and (b), illustrate
different models considered for the ionizing continuum.  Model (a) is an
extrapolation of the EUV continuum throughout the X-ray band.  Model (b)
connects the EUV continuum to the ROSAT data by introducing a break
at the location of the He II ionization edge (54 eV).}
\label{fig:broadband}
\end{figure}

To estimate the soft X-ray flux from the {\it ROSAT} PSPC data, 
we use the total count rate of
$7.02 \times 10^{-3} \ {\rm count \ s}^{-1}$ in the observed energy band 
$0.1-2.5 \ {\rm keV}$ ($0.3-8.4 \ {\rm keV}$ in the rest frame)
and the Galactic neutral hydrogen column 
density of $N(H) = 1.30 \times 10^{20} \ {\rm cm}^{-2}$.  
Assuming a power-law index of $\alpha_x=-1.7$,
a typical value for the soft X-ray ($\sim 0.1-{\rm a \ few \ keV}$)
continuum of radio-quiet QSOs as derived from
PSPC observations (Schartel {\it et al.} 1995\markcite{scea95}; 
Laor {\it et al.}\ 1997\markcite{laea97}),
the observed flux at 1 keV is $5.7 \times 10^{-32}$ \nuflux, which corresponds
to a flux of $1.3 \times 10^{-31}$ \nuflux\ at
1 keV in the QSO rest frame.  This is an order of magnitude less flux
than expected based on an extrapolation of the extreme-UV
continuum.  We illustrate this in Figure~\ref{fig:broadband} by
plotting the extrapolated extreme-UV continuum and the derived X-ray
continuum.  We also show the continuum derived from the {\it ROSAT} data
assuming $\alpha _x =-1.0$, a more typical value for harder X-rays
($\gtrsim {\rm a \ few \ keV}$; 
Comastri {\it et al.}\ 1992\markcite{coea92};
Williams {\it et al.}\ 1992\markcite{wiea92}).

The relative luminosity between
the optical and the X-ray is commonly represented by the parameter 
$\alpha_{ox}=(\log F_o - \log F_x)/2.605$, where $F_o$ is the spectral flux at 
2500 \AA\ and $F_x$ is the flux at 2 keV.  An extrapolation of the near-UV
continuum yields a rest-frame flux at 2500 \AA\ of 
$F_o = 2.8 \times 10^{-27}$ \nuflux, and from the X-ray data we
infer a rest-frame flux at 2 keV of 
$F_x = 4.0 \times 10^{-32}$ \nuflux, yielding 
$\alpha_{ox}=1.86$.  Such a high value for $\alpha_{ox}$ is consistent with 
the results of Green \& Mathur (1996\markcite{grma96}), who find values of 
$\alpha_{ox} \gtrsim 1.8$ in a sample of BALQSOs observed with the {\it ROSAT}
PSPC.  The authors attribute the lack of X-ray flux to absorption by the 
BAL gas and derive BAL column densities 
$\gtrsim 2 \times 10^{22} \ {\rm cm}^{-2}$.  It is therefore reasonable to
suspect that the X-ray continuum of SBS1542+541 may be absorbed by the BAL gas.
Hence, to choose an intrinsic ionizing continuum for our photoionization 
models, we defer to the UV observations and use the extrapolated
power law with index $\alpha=-1.59$ shown in Figure~\ref{fig:broadband}.  
The possibility of X-ray absorption is further discussed in \S\ref{sec:xray}.
In \S\ref{sec:altcont} we discuss the effect
of choosing an alternative continuum shape. 

All of our calculations were performed with the photoionization code 
CLOUDY (version 90.01; Ferland 1996\markcite{fer96}).
We assume that the BAL gas is in photoionization equilibrium 
with the intrinsic QSO continuum and has solar chemical abundances.
We use a constant density of 
$n_H=10^8 \ {\rm cm^{-3}}$, although it has been previously established 
({\it e.g.}\ Hamann {\it et al.}\ 1995\markcite{haea95}) 
in similar models that the ionization state of the gas is not 
sensitive to the absolute density (up to a limit of 
$\sim 10^{11}-10^{12} \ {\rm cm^{-3}}$).  Hence our results are valid for any 
reasonable space density, so long as it is constant.
The continuum normalization is determined by the ionization parameter $U$, 
defined as the ratio of the number density of ionizing photons 
(energy above the ionization threshold of hydrogen) to hydrogen nuclei,
\begin{equation}
U=\int_{\nu_{th}}^\infty \frac{F_\nu}{ch\nu n_H} d\nu,
\end{equation}
where $\nu_{th}$ is the frequency at the hydrogen ionization threshold.
We do not assume that the gas is optically thin to the ionizing continuum.  In
fact, given the necessarily large total column densities that we infer, 
continuum opacity must be an important factor and will alter the spectral
energy distribution of the ionizing continuum as it propagates through the BAL
gas, resulting in changes in the ionization state.  

\subsubsection{Single-Slab Model}

\begin{figure}
\plotland{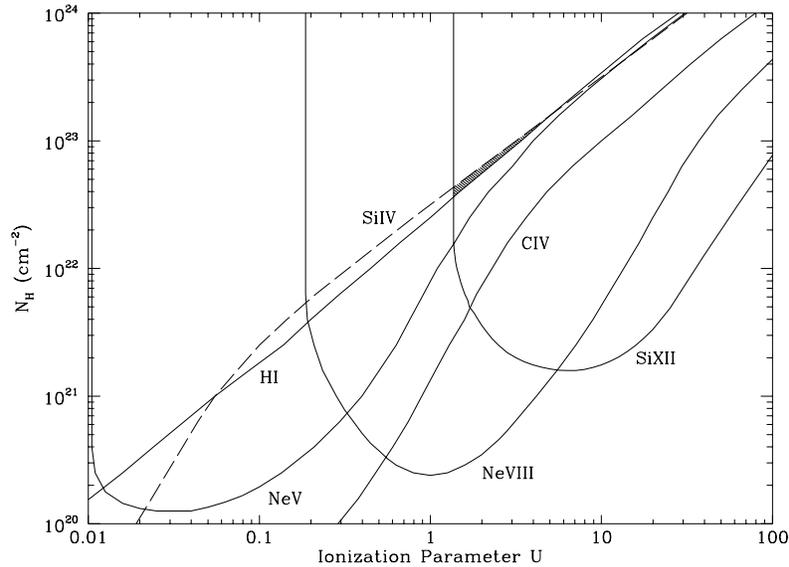}
\caption{\protect\captionsize \protect\captionbaseline
Constraints on the total column density and
ionization parameter for a single-slab model.  The solid lines are contours of 
constant ionic column density corresponding to our measured lower limits, 
while the dashed line represents an upper limit.
A particular choice of parameters must lie in the parameter space above the 
solid lines and below the dashed line to produce a model consistent with the 
column density limits.  The shaded region satisfies all of these 
constraints.}
\label{fig:nuplaneone}
\end{figure}

The simplest model of the BAL gas is a single slab characterized by
its total column density $N_H$ and the ionization parameter $U$
at the face of the slab.  We use our measured limits on the ionic column 
densities to constrain the
total column density and ionization parameter for such a model.  We illustrate 
this in Figure~\ref{fig:nuplaneone} by plotting in the $N_H-U$ parameter space 
contours of constant ionic column density equal to a few of our lower limits, 
including those which provide the most restrictive constraints on the 
parameters.  The contour for \ion{H}{1} corresponds to a column density of 
$10^{16.8}$ \col.  Models consistent with our lower limits on the 
column densities must lie above the solid lines.
We also include a constraint from the lack of \ion{Si}{4} absorption.
By fitting the data assuming a flat absorption profile with a centroid at 
$-11400$ \kms, a width of $2000$ \kms, and a covered fraction of $0.2$, we 
derive an upper limit on the 
column density of \ion{Si}{4} of $10^{14.1}$ \col\ with 90\% confidence.
This upper limit is represented by the dashed line in 
Figure~\ref{fig:nuplaneone}.  Models consistent with the upper limit must lie 
below the dashed line to avoid producing an observable amount of \ion{Si}{4}.  
Thus the shading indicates the region of parameter space consistent with our 
column density limits in this simple model.

The existence of a region in the $N_H-U$ parameter space that satisfies 
all of our column density constraints means that, based on the column 
density limits alone, we could explain the observations by a model of a 
single, marginally optically thick, BAL region of constant density.  In such 
a model the large range of ionization states is possible because the very 
high-ionization states would exist near the face of the slab, whereas the 
lower-ionization states exist deeper in the slab ({\it i.e.}\ farther from the 
continuum source) as a result of X-ray self-shielding.  However, to 
understand the BAL system we would like to be able to explain not only the
ionization state of the gas but also the covered fractions, and in particular 
the correlation of a higher covered fraction with a higher state of 
ionization.  This would be impossible to achieve in a single-zone model.  We 
now discuss how we can modify this simple single-slab picture to arrive at a 
more plausible explanation of the covered fractions.

\subsubsection{Two-Slab Model\label{sec:twoslab}}

\begin{figure}
\epsscale{0.7}
\plotone{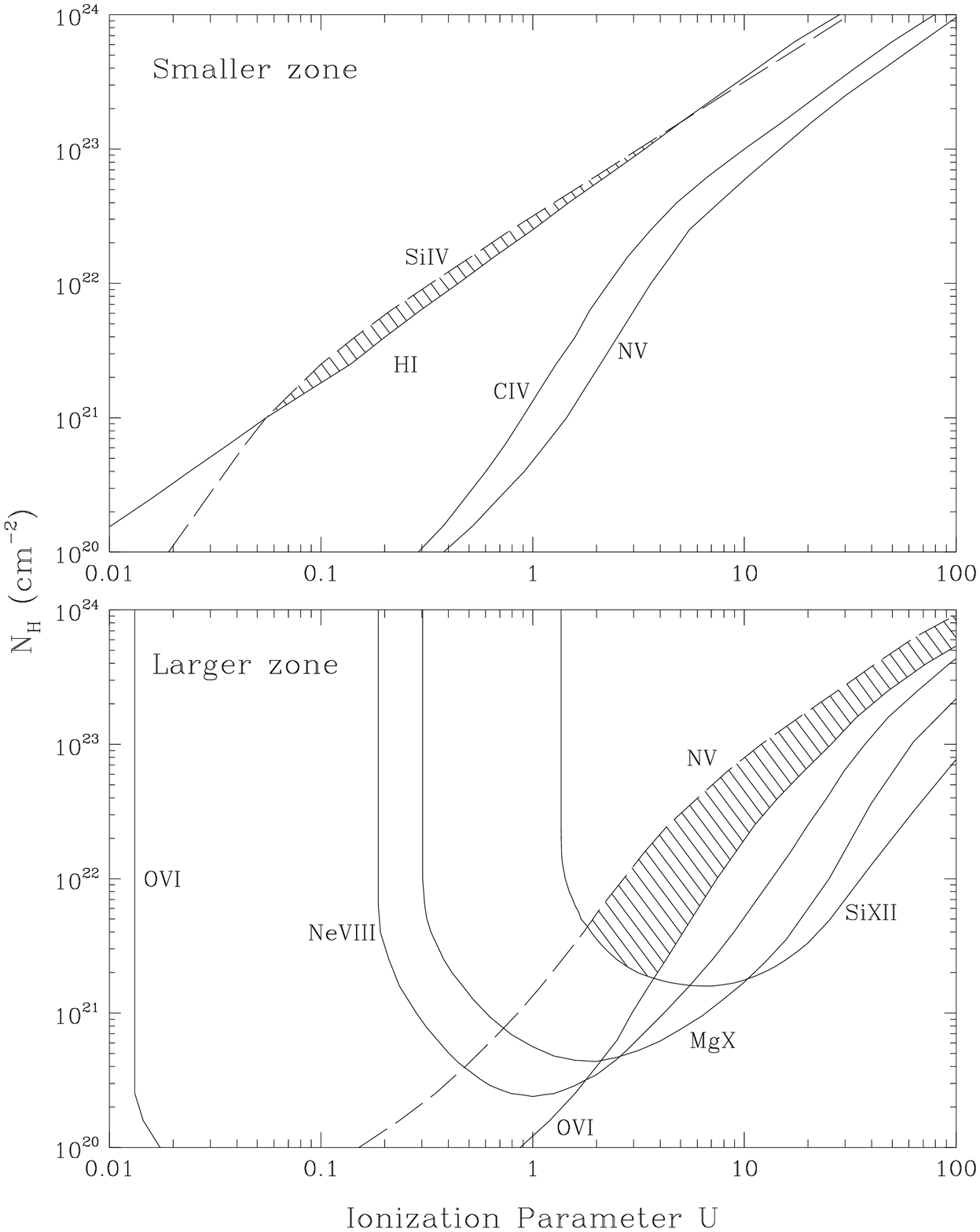}
\caption{\protect\captionsize \protect\captionbaseline
Constraints on the total column density and
ionization parameter for a two-slab model.  The system is separated into 
a larger zone and a smaller zone based on the observed covered fractions.  
The solid lines indicate lower limits on the ionic column densities, 
the dashed lines indicate upper limits, and the shaded regions satisfy the 
column density constraints applied to each zone.}
\label{fig:nuplanetwo}
\end{figure}

Multiple zones are necessary to explain the range in covered fractions
that we observe.  Empirically, what is required is that the high-ionization
gas have a larger covered fraction than the lower-ionization gas.
Here we divide the BAL region into two
zones, one with a low covered fraction contributing the absorption from the 
higher-order Lyman lines and perhaps also \ion{C}{4} and \ion{N}{5}, and 
another with a higher covered fraction providing the dominant absorption 
source of the higher-ionization species, up through \ion{Si}{12}.  We will 
refer to these as the ``smaller'' and ``larger'' zones, respectively.
Figure~\ref{fig:nuplanetwo} shows the constraints for each zone provided by 
our observed column density limits as in the single-slab case.
For the larger zone we apply an additional constraint provided by
the fact that \ion{N}{5} absorption clearly cannot be optically thick in this 
zone.  In the same manner as described in the previous section for \ion{Si}{4},
but here assuming a covered fraction of $0.4$, we place an upper limit of 
$10^{15.2}$ \col\ on the column density of \ion{N}{5}.  Placing 
upper limits on the column densities of \ion{C}{4} and \ion{O}{5} 
results in constraints that are very similar but slightly less-restrictive, so 
they are not shown.

Further plausible physical assumptions can place more restrictions on the
possible range of parameter space covered by the two zones.
For example, we could assume that the clouds that make up the BAL gas are
equidistant from the central radiation source, but have small, dense cores
surrounded by lower-density halos.  The cores would naturally have a lower
ionization parameter, and they would cover a smaller fraction of the source.
More generally, one could envision a density-fluctuated medium containing
small, dense clouds of radius $R_s$, and larger, lower-density clouds of
radius $R_l$ satisfying the constraint $R_s < R_l$, where we use the subscripts
$s$ and $l$ to denote the smaller and larger zones, respectively.  We can 
relate the cloud radii to their column densities as $R \propto N_H / n_H$.
Since $U \propto 1 / (r^2 n_H)$, we have $R \propto r^2 N_H U$.
Assuming the two zones are at roughly the same 
distance $r$ from the continuum source, we can therefore write the size 
constraint as $N_{H,s} U_s < N_{H,l} U_l$.
This reduces the range of acceptable two-slab models in the four-dimensional 
parameter space spanned by $U_s$, $N_{H,s}$, $U_l$, and $N_{H,l}$. 
By considering the effect of this constraint on the solutions shown in
Figure~\ref{fig:nuplanetwo}, we conclude that, in this density fluctuation
model, the ionization parameter in the 
smaller zone must be a factor $\gtrsim 2$ lower than in the larger zone,
which would correspond to a larger density in the smaller zone by the same 
factor.

Density fluctuations in the BAL gas and the resulting correlation
between covered fraction and ionization state could have implications for the 
determination of chemical abundances in BAL systems.  Studies of BAL systems
typically find that a large range in ionization
parameters (more than an order of magnitude) and large abundance modifications
relative to solar (as much as an order of magnitude, or even two for 
nitrogen) are necessary to explain the observed column densities
({\it e.g.}\ Turnshek {\it et al.}\ 1996\markcite{tuea96}).  The large 
metallicity
enhancements are required primarily because of the observed weakness of 
the \lya\ BAL, which is used to derive the \ion{H}{1} column density.
However, in the BAL system of SBS1542+541, the small equivalent width
of \lya\ despite a large optical depth ($\tau_{Ly\alpha} > 50$)
is explained by
the fact that most of the \ion{H}{1} covers only a small fraction
of the continuum source.  As a result, the contribution from most
of the \ion{H}{1} is ``hidden'' beneath the absorption from more
highly-ionized gas with a much lower column density but higher covered
fraction.  This situation is revealed by the direct observation of the
saturated Lyman lines of order Ly$\gamma$ and higher, lines that in a more
typical BAL system would not be visible due
to blending with other lines, particularly \ion{C}{3} \lam 977.  If
inhomogeneity of the BAL gas is a generic feature of BAL systems, 
particularly that which would result in large columns of hidden \ion{H}{1},
then this could affect (and make more difficult) the measurement 
and interpretation of BAL column densities and chemical abundances.

\begin{figure}
\plotland{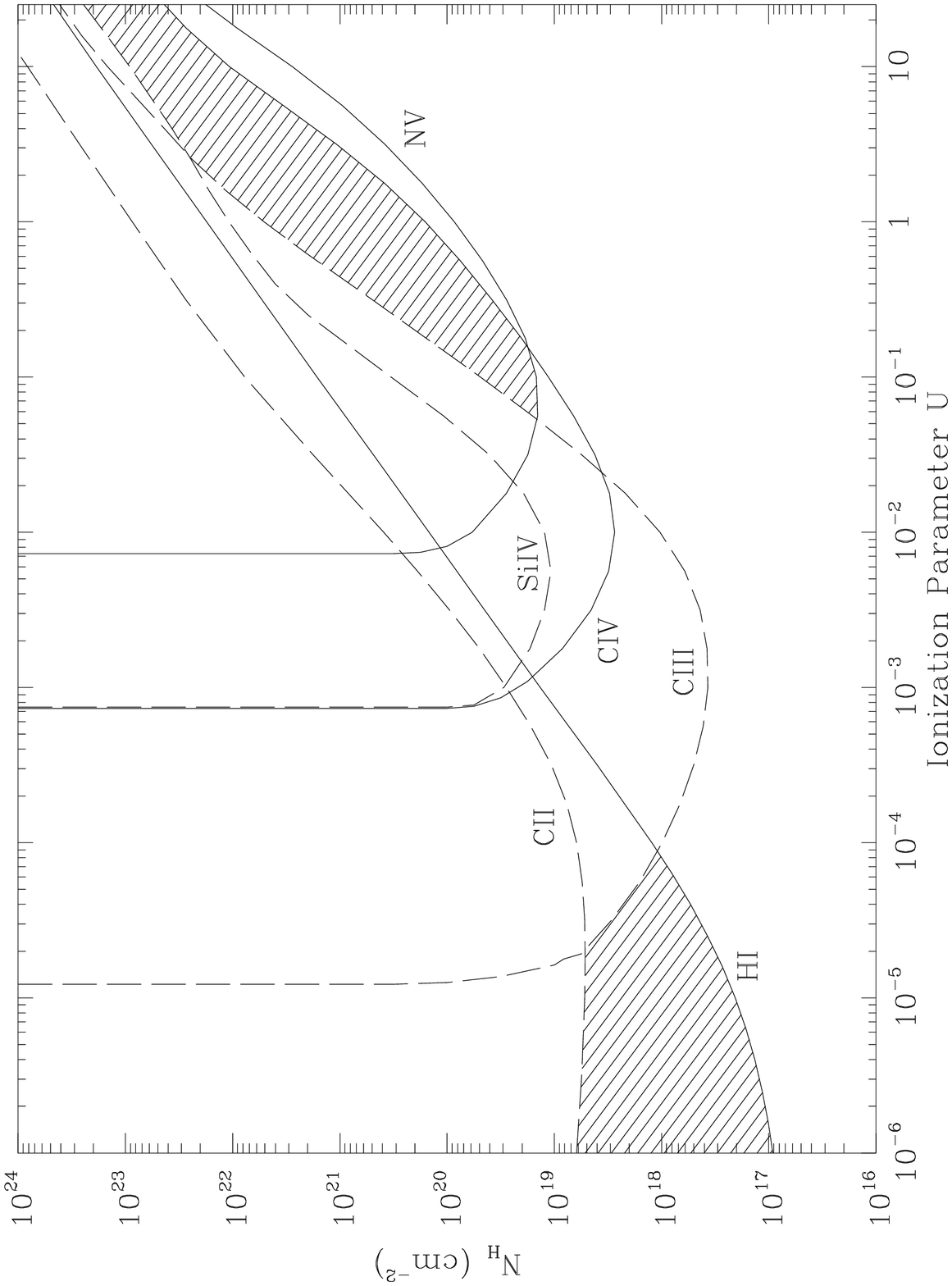}
\caption{\protect\captionsize \protect\captionbaseline
Some contours of constraint for the ions with
a small covered fraction as shown in the top of Figure~11 but now produced 
using continuum shape (b) from Figure~9.  At least two zones, represented by 
the shaded
regions, are necessary to explain the observations: a low column density, 
weakly-ionized zone (left) producing the H I absorption and a more 
highly-ionized, larger column density zone (right) producing the C IV and N V
absorption.  Upper limits on some low-ionization species
constrain the zones.}
\label{fig:nuplanealt}
\end{figure}

In this model, the source of the \ion{H}{1} absorption is constrained to the 
small region of parameter space shown in Figure~\ref{fig:nuplanetwo} by the
tight upper limit on \ion{Si}{4} absorption.  However, it is 
possible that the \ion{H}{1} absorption arises in a very weakly-ionized 
zone ($U \lesssim 10^{-4}$) with a low enough column density 
($N_H \lesssim 10^{18}-10^{19}$ \col) that no strong low-ionization metal 
absorption lines are produced.  If such a zone exists, 
then the \ion{C}{4} and \ion{N}{5}
lines require an additional higher-ionization zone.  This arrangement
is illustrated in Figure~\ref{fig:nuplanealt}, as calculated using
continuum shape (b) from Figure~\ref{fig:broadband} 
(see \S\ref{sec:altcont}).  Alternatively,
they could be produced in the larger zone if the lines are assumed to be not 
optically thick.  This is unlikely, however, since the profiles, 
particularly \ion{C}{4}, strongly suggest saturation.

Although a two-zone model is the simplest one that can explain the basic 
properties of the system, including the observed line strengths and the 
general trend of the covered fractions, there are complications.  
In particular, the observed
range in covered fractions for the highly-ionized species, especially the
apparently high covered fraction of \ion{Ne}{8},
cannot be explained by a simple model.  If this is a real effect, 
then the physical structure
of the BAL gas could be extremely complex, and much more detailed
modeling would be necessary to reproduce this peculiar behavior.  However,
it is entirely possible that this effect is simply due to the uncertainties
induced by blending and the choice of fitting profile.

\subsubsection{Disk-Like Geometry\label{sec:disk}}

\begin{figure}
\epsscale{0.7}
\plotone{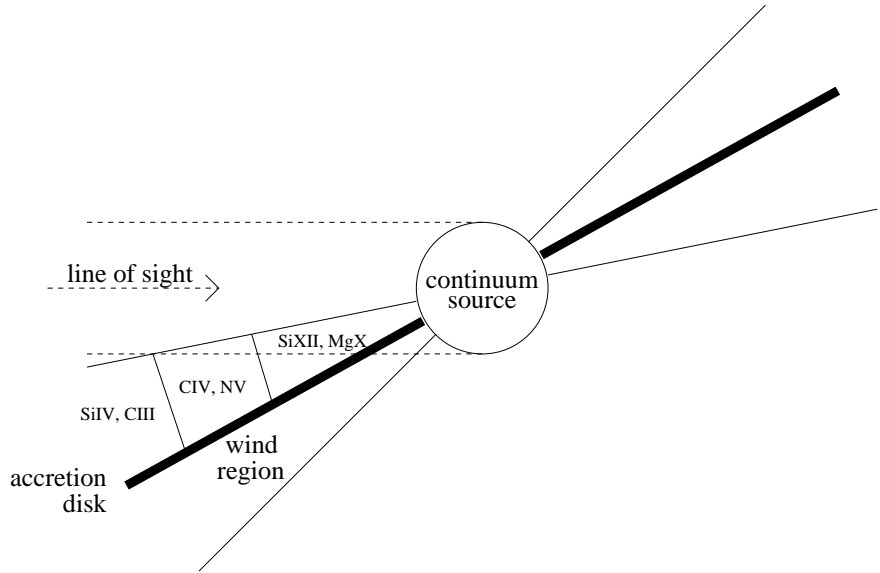}
\caption{\protect\captionsize \protect\captionbaseline
Schematic illustration of a disk-like BAL geometry
as viewed from a line of sight just above the disk.  Because of the geometry, 
the more weakly-ionized gas, which lies farther away from the continuum source,
will cover less of the source along such a line of sight.  A typical BAL sight 
line would be directly into the disk wind.}
\label{fig:disk}
\end{figure}

The correlation of covered fraction with ionization state can also be 
explained by more complex physical models.  It is plausible that the BAL 
region has an overall disk-like geometry as suggested by the high continuum
and even higher trough polarization observed in BALQSOs 
(Goodrich \& Miller 1995\markcite{gomi95}; 
Cohen {\it et al.}\ 1995\markcite{coea95}).
This has been explored theoretically in the disk-wind models of AGN broad line 
regions
(Murray {\it et al.}\ 1995\markcite{muea95}; 
Murray \& Chiang 1995\markcite{much95}).
One can imagine that we are looking along a particular line of sight, just
above (or below) the disk, as illustrated schematically in 
Figure~\ref{fig:disk}.  The 
states of lower ionization, being farther from the continuum source, would 
naturally have a smaller covered fraction.  An appealing consequence of this 
interpretation is that the BAL region that we observe in SBS1542+541 would 
not be required to be inherently exceptional in its lack of absorption from 
states of lower ionization such as \ion{Si}{4}, \ion{O}{4}, and \ion{C}{3}.  
These ions could be part of the BAL region but be far enough from the 
continuum source that they lie outside our line of sight.

\subsection{Possible X-Ray Absorption\label{sec:xray}}

The existence of high column densities of highly-ionized gas suggests that
the UV absorption system of SBS1542+541 is associated with a
``warm absorber'' as seen in the X-ray spectra of many low-redshift AGN.  
Models of X-ray warm absorbers (Netzer 1993\markcite{net93}; Krolik \& Kriss 
1995\markcite{krkr95}; Netzer 1997\markcite{net97}) predict that 
the strongest features should be the K ionization edges of \ion{O}{7} and 
\ion{O}{8} as well as K and L edges of ionization stages of iron 
$\sim$\ion{Fe}{20}.  According to our photoionization calculations, given 
the likely range of ionization parameters for the BAL gas in SBS1542+541, 
the dominant stages of oxygen should be 
\ion{O}{7} and \ion{O}{8}.  The predicted large column densities of these ions
should produce observable ionization edges at $0.22$ keV (\ion{O}{7}) and 
$0.26$ keV (\ion{O}{8}), each with an optical depth of $\tau > 0.2$.  
Observations of these X-ray features would enable us to form a 
much more complete picture of the ionization state of the system, making 
this object a good candidate for further study of the unification 
of X-ray/UV absorbers (Mathur {\it et al.}\ 1994\markcite{mwef94};
Mathur, Elvis, \& Wilkes 1995\markcite{mew95}; 
Mathur, Wilkes, \& Aldcroft 1997\markcite{mwa97}).

\subsection{Continuum Shape\label{sec:altcont}}

We now consider the possibility that the weak X-ray
flux is an intrinsic characteristic of the continuum being intercepted
by the BAL region; {\it i.e.} it is not due primarily to X-ray absorption by 
the BAL gas itself.  For this purpose we have repeated our photoionization
calculations using the ionizing continuum shape labeled
(b) in Figure~\ref{fig:broadband}.  Here we connect the
X-ray continuum inferred from the {\it ROSAT} flux assuming $\alpha_x = -1.7$
with the EUV continuum by arbitrarily introducing a break in the EUV 
power-law index
at the location of the \ion{He}{2} ionization edge (54 eV).  The break point 
is chosen to be consistent with the UV observations (below our
wavelength coverage), but at a low enough energy that it should maximize
the effects on the results.  

Naturally, due to the softer continuum,
much higher ionization parameters are necessary to produce the highest
observed ionization states, $U>10$ for \ion{Si}{12}.  More interestingly,
the contours of constraint from \ion{H}{1} and \ion{Si}{4} shift relative
to one another such that the solution for the smaller zone shown in
Figure~\ref{fig:nuplanetwo} does not exist for this model.  That is, one
cannot construct a simple model of the smaller zone that produces the
necessary columns of \ion{H}{1}, \ion{C}{4}, and \ion{N}{5} that does
not also produce a large column density of \ion{Si}{4}.  The only 
explanation in this model for the \ion{H}{1} absorption, therefore,
is that there is a very weakly-ionized, low column density zone as mentioned 
in \S\ref{sec:twoslab}.  Figure~\ref{fig:nuplanealt} shows the 
location of such a zone in the $N_H-U$ parameter space, using upper limits
of $10^{15.2}$ \col\ on \ion{C}{2} (from \lam 1335) and $10^{14.0}$ \col\
on \ion{C}{3} (from \lam 977), again derived in the same manner as for
\ion{Si}{4}.
This would require an enormous spread in ionization parameters (5 orders of
magnitude) between the different zones.  Since the existence of a zone
with such a different ionization state, yet having a very similar distribution 
in velocity space, is not particularly appealing, this indirectly 
suggests that the ionizing continuum extrapolated from the EUV
(labeled (a) in Figure~\ref{fig:broadband}) is a
better representation of the true continuum and that the weak X-ray
flux is indeed a consequence of X-ray absorption.

\section{SUMMARY\label{sec:summary}}

The BAL system of SBS1542+541 has several unusual properties.  The most 
interesting are:
\begin{enumerate}
\item{The state of ionization is extremely high, higher than what
is typically inferred for ``classical'' BAL systems from the study of the UV 
absorption lines.
Strong absorption is present from the highly-ionized lithium-like ions 
of \ion{O}{6}, \ion{Ne}{8}, \ion{Mg}{10} and even \ion{Si}{12}, which
has a creation ionization potential of 476 eV.  Absorption from
\ion{Si}{4}, nearly always present in BAL spectra, is undetected.
It should be noted that high ionization similar to this has been
suggested in connection with BALs, but this is based on X-ray
rather than UV absorption
properties (Mathur, Elvis, \& Singh 1995\markcite{mes95}).  The high state of 
ionization and large column density ($>2 \times 10^{21}$ \col) 
suggest that the BAL system should be associated with an X-ray 
``warm absorber''.}
\item{The covered fractions for the various absorbing ions are in at least
some cases significantly less than unity.  Even more interesting is that the
covered fraction is apparently ion-dependent with a general trend towards 
higher covered fractions with higher states of ionization.  This could be 
caused by either a special line of sight through a BAL disk or by 
the existence of multiple ionization zones in the BAL gas, possibly as a 
result of density fluctuations.}
\end{enumerate}
These properties make the BAL system of SBS1542+541 observationally 
unique and certainly intriguing.  It begs the question---is this 
system extraordinary, or is it telling us something important about the
character of BALQSOs in general?  The weakness of the standard BALs of
\lya, \ion{C}{4}, \ion{N}{5}, and \ion{Si}{4} is clearly unusual compared to
``classical'' BALQSOs.  However, the lack of observed \ion{Si}{4} and other
low-ionization species itself is perhaps not so surprising given the
weakness of \ion{C}{4}, particularly if complex covered-fraction effects
are involved.  The very high-ionization species we observe in SBS1542+541
occur in the rest-frame EUV, a spectral region not usually observable
due to intervening Lyman-limit absorbers.  Hence the existence of these ions is
not inconsistent with other BALQSOs, for which it is not possible
to confirm or deny their presence.  It therefore seems plausible that the
ionization state of this system is similar to that of normal BALQSOs.
If the existence of large column densities of very highly-ionized gas 
and/or non-trivial ionization structure as we find in SBS1542+541 are common 
properties of BAL systems, then this object could indeed be providing
important clues to the nature of the BAL phenomenon.

\acknowledgments

This work is based on observations with the NASA/ESA
Hubble Space Telescope, obtained at the Space Telescope Science
Institute, which is operated by the Association of Universities for
Research in Astronomy, Inc., under NASA contract NAS5-26555.
Support for this research was provided by NASA grant NAG5-1630 for the
Faint Object Spectrograph team,
and NASA contract NAS5-27000 for the Hopkins Ultraviolet Telescope.

\end{document}